%% file: polarMAP030917.tex
\documentclass[peerreview]{IEEEtran}
\usepackage{url}
\usepackage{graphicx}
\usepackage{amsmath}
\usepackage{amssymb}
\usepackage{mathtools}
\usepackage{dsfont}
\mathtoolsset{showonlyrefs=true}
\usepackage{algorithm}
\usepackage[noend]{algpseudocode}
\usepackage{color}

\allowdisplaybreaks

\makeatletter
\providecommand*{\boxast}{%
	\mathbin{
		\mathpalette\@boxit{*}%
	}%
}
\newcommand*{\@boxit}[2]{%
	\sbox0{$\m@th#1\Box$}%
	\ifx#1\displaystyle \ht0=\dimexpr\ht0+.05ex\relax \fi
	\ifx#1\textstyle \ht0=\dimexpr\ht0+.05ex\relax \fi
	\ifx#1\scriptstyle \ht0=\dimexpr\ht0+.04ex\relax \fi
	\ifx#1\scriptscriptstyle \ht0=\dimexpr\ht0+.065ex\relax \fi
	\sbox2{$#1\vcenter{}$}
	\rlap{%
		\hbox to \wd0{%
			\hfill
			\raisebox{%
				\dimexpr.5\dimexpr\ht0+\dp0\relax-\ht2\relax
			}{$\m@th#1#2$}%
			\hfill
		}%
	}%
	\Box
}
\makeatother

\input{shortcuts.tex}

\begin{document}

\title{Performance Bounds of Concatenated Polar Coding Schemes}

\author{Dina Goldin and David~Burshtein,~\IEEEmembership{Senior Member,~IEEE}
\thanks{This research was supported by the Israel Science Foundation, grant no. 1082/13.}
\thanks{D.\ Goldin is with the school of Electrical Engineering, Tel-Aviv University, Tel-Aviv 6997801, Israel (email: dinagold@post.tau.ac.il).}
\thanks{D.\ Burshtein is with the school of Electrical Engineering, Tel-Aviv University, Tel-Aviv 6997801, Israel (email: burstyn@eng.tau.ac.il).}
\thanks{This material in this paper was presented in part in ISIT 2017.}
}

\markboth{Submitted to IEEE Transactions on Information Theory}{Goldin and Burshtein: Performance Bounds of Concatenated Polar Coding Schemes}

\maketitle \setcounter{page}{1}

\begin{abstract}
A concatenated coding scheme over binary memoryless symmetric (BMS) channels using a polarization transformation followed by outer sub-codes is analyzed. Achievable error exponents and upper bounds on the error rate are derived. The first bound is obtained using outer codes which are typical linear codes from the ensemble of parity check matrices whose elements are chosen independently and uniformly. As a byproduct of this bound, it determines the required rate split of the total rate to the rates of the outer codes. A lower bound on the error exponent that holds for all BMS channels with a given capacity is then derived. Improved bounds and approximations for finite blocklength codes using channel dispersions (normal approximation), as well as converse and approximate converse results, are also obtained. The bounds are compared to actual simulation results from the literature.
For the cases considered, when transmitting over the binary input additive white Gaussian noise channel, there was only a small gap between the normal approximation prediction and the actual error rate of concatenated BCH-polar codes.
\end{abstract}

\section{Introduction} \label{sec:intro}
Polar coding, introduced by Arikan~\cite{arikan2009channel}, is an exciting development in coding theory. Arikan showed that, for a sufficiently large blocklength, polar codes can be used for reliable communications at rates arbitrarily close to the symmetric capacity 
of an arbitrary binary-input channel.
Various decoding algorithms that improve Arikan's successive cancellation (SC) 
decoder were shown since then. A notable example is list SC decoding 
\cite{tal2015list} with the possible incorporation of CRC bits.
Various architectures have been considered for parallel efficient implementation of SC and list SC decoding with improved throughput, e.g. \cite{leroux2013semi,li2014low,yuan2015low,xiong2015symbol}. Those architectures involve 
decomposing the overall polar code into an inner code and an outer code, and using SC to decode the inner code and maximum-likelihood (ML) to decode the outer code.

\begin{figure} 
\centering
\includegraphics[width=0.5\columnwidth]{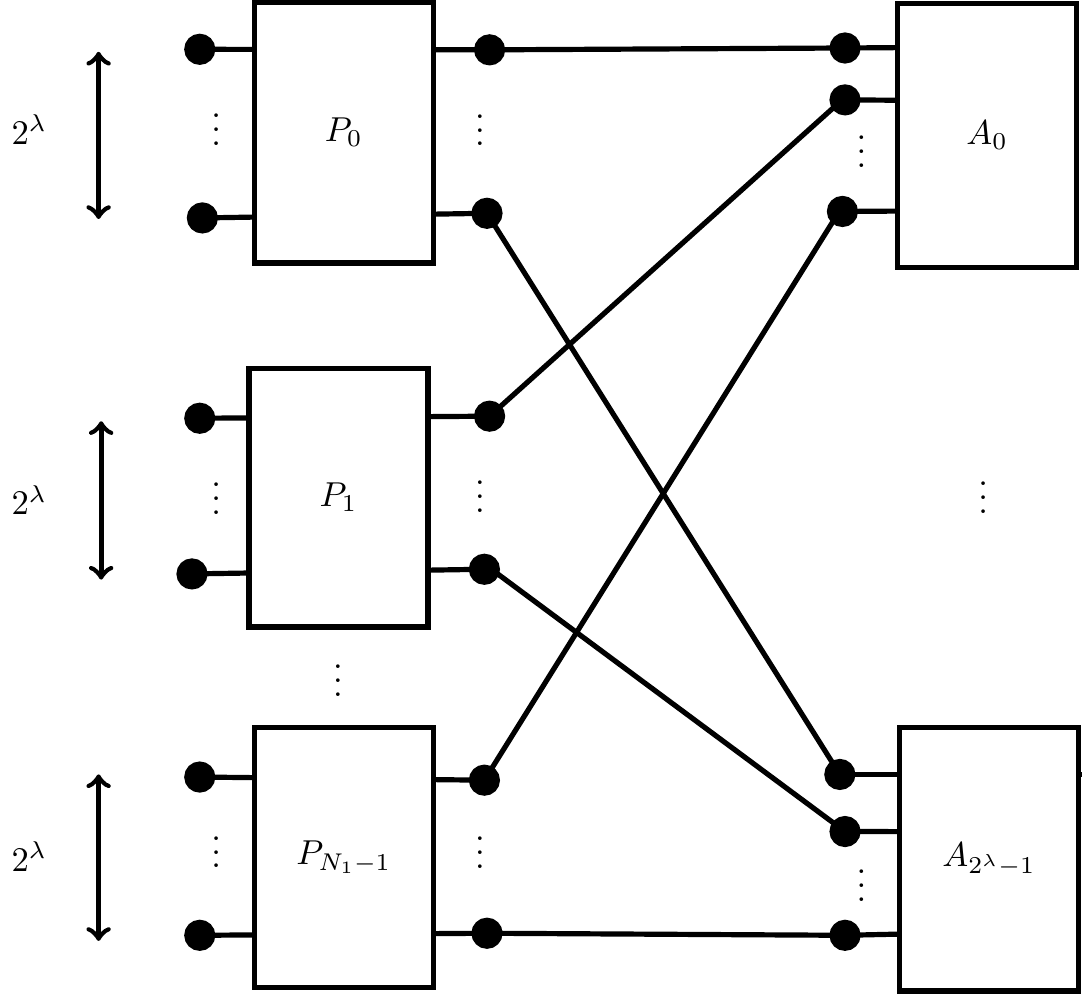}
\vspace{-10pt}
\caption{Concatenation of $N_1$ polar codes of length $2^{\lambda}$ with $2^{\lambda}$ outer codes of length $N_1$}
\label{fig:concat}
\end{figure}

Using other outer codes such as powerful algebraic codes with approximated ML decoding is also possible. Interleaved concatenation of inner polar codes with outer Reed-Solomon (RS) and Bose Chaudhuri and Hocquenghem (BCH) outer codes was studied in \cite{mahdavifar2014performance} and \cite{trifonov2011generalized}, respectively, and further studied in \cite{wang2016interleaved}, that also proposed using convolution outer codes.
In this work, we study the interleaved concatenated scheme of polar codes with good outer codes \cite{mahdavifar2014performance,trifonov2011generalized,wang2016interleaved}.
This scheme is described in Fig. \ref{fig:concat}. The encoding is performed from right to left as follows. First, we use $2^{\lambda}$ outer codes with rates $R_i$, $i=0,\ldots,2^\lambda-1$, to encode the information bits, creating $2^\lambda$ codewords, each of length $N_1$. The resulting codewords are interleaved and processed by $N_1$ polar encoders of length $2^{\lambda}$ as shown in Fig. \ref{fig:concat}. We obtain a code with blocklength $N=N_1\cdot 2^{\lambda}$ and rate $R=\sum_{i=0}^{2^{\lambda}-1}R_i/2^{\lambda}$.
	
The decoding can also be described using Fig. \ref{fig:concat}. However it is performed from left to right. As described in \cite{wang2016interleaved}, the first information bit of each of the $N_1$ polar codes is decoded in parallel, using a soft-decision algorithm that produces log-likelihood ratio (LLR) values. These LLRs are used as the input of the decoder of the code $A_0$, and the decisions of that decoder are passed back to the polar decoders, and used in calculating the LLR of the second information bit of each polar code. These LLRs are then used as the input of the decoder of $A_1$, etc. In general, when the LLRs of the $i$-th information bits of the $N_1$ polar codes are calculated, the previous $i-1$ information bits of these codes are available from decoding $A_0,A_1,\dots,A_{i-2}$.
We described this coding scheme as a combination of block SC decoding and optimal (ML), or non-optimal decoding of sub-codes with blocklength $N_1$ in \cite{goldin2016block}. If the outer codes are polar as well, this is the scheme proposed in \cite{li2014low} for efficient parallel decoding of polar codes. If the outer codes are Reed-Solomon, BCH or convolutional codes, this is the scheme studied in \cite{mahdavifar2014performance,trifonov2011generalized,wang2016interleaved}. Note that unlike standard polar codes, in the concatenated polar coding scheme there are no frozen bits, unless $R_i=0$, when the whole block $A_i$ is frozen. We also note that a close to ML, computationally efficient decoding of short outer codes can be realized using various algorithms such as the ordered statistics decoder (OSD) \cite{fossorier1995soft}, the box-and-match decoder \cite{valembois2004box}, used in~\cite{trifonov2011generalized}, or the recent machine learning-based schemes presented in~\cite{nachmani2016learning,nachmani2017deep,tenbrink,cammerer2017scaling} and references therein, that may be efficient for small blocklength codes (especially when using hardware implementations). The decoder of the concatenated coding scheme may possess an improved throughput compared to list SC decoding of polar codes, provided the outer codes can be decoded efficiently.

In this work we analyze the performance of the concatenated polar coding scheme. Our main interest is in the case where $\lambda$ is small (e.g., $\lambda=1,2,3$), and the blocklength of each outer code, $N_1$, is also relatively small (e.g., of order 100) such that the OSD algorithm or the other methods mentioned above, can be used to decode the outer codes with a reasonable computational effort. As a motivating example, suppose we need to design an error correcting code with blocklength which is about $N=256$, and rate which is about $1/2$. As will be described in detail later in the paper, in order to closely approach the error rate of the best code under these conditions, as predicted by the normal approximation in \cite{polyanskiy2010channel}, one may apply the OSD algorithm to a BCH code with blocklength $255$ and rate $R = 131/255 \approx 1/2$. However, the resulting computational complexity would be prohibitive for actual, real time, low cost and low power applications. Alternatively, as will be shown in Section~\ref{sec:simulation}, one may construct a BCH-polar concatenated scheme with $\lambda=1$ and $N_1=128$. The total blocklength is $N=256$, and the total rate is $R=1/2$. The first outer code is a BCH code with rate $R_1=36/128$. The second outer code is a BCH code with rate $R_2=92/128$ (these rates were determined from our analysis in Section \ref{sec:improved_bnds_and_approx}). In order to decode the scheme, one needs to apply the OSD algorithm twice, first to the lower rate code, and then to the higher rate code. As Fig. \ref{fig:polarBCH128time} shows, the computational complexity of the decoder of the first scheme (BCH with blocklength $255$) is larger by about three orders of magnitude compared to the BCH-polar concatenated scheme. On the other hand, as Fig. \ref{fig:polarBCH} shows, the resulting performance degradation, when using the BCH-polar concatenated code, for frame error rate of $10^{-3}$, is only about $0.6{\rm dB}$ in the signal to noise ratio. Fig. \ref{fig:polarBCH} also shows that our prediction to the performance of the BCH-polar concatenated scheme is very accurate.

In Section \ref{sec:background} we provide some brief background on polar codes, and fix the notations. In Section \ref{sec:bounds_on_error_exps} we obtain achievable error exponents and upper bounds on the error rate for the concatenated polar coding scheme. Our first lower bound on the error exponent (upper bound on the error rate) can be achieved using outer codes which are typical linear codes from the ensemble of parity check matrices whose elements are chosen independently and uniformly (i.e., they are set to $\{0,1\}$ with probabilities $\{1/2,1/2\}$). As a byproduct of this bound, we determine the required rate split of the total rate to the rates of the outer codes. We then obtain a lower bound on the error exponent that holds for all binary memoryless symmetric (BMS) channels with capacity $I$.
In Section \ref{sec:improved_bnds_and_approx} we derive improved bounds and approximations using channel dispersions (normal approximation) for finite blocklength codes. We also derive converse and approximated converse results. In Section \ref{sec:simulation} we compare our bounds to actual simulation results.
For the cases considered, when transmitting over the binary input additive white Gaussian noise channel (BIAWGNC), there was only a small gap between the normal approximation prediction and the actual error rate of concatenated BCH-polar codes. Section \ref{sec:conclusion} concludes the paper. 
\section{Background on Polar Codes} \label{sec:background}
Consider a BMS $W:\cX\rightarrow\cY$ with input alphabet $\cX=\left\{0,1\right\}$ and output alphabet\footnote{The assumption that the channel is discrete is made for notational convenience only. For continuous output channels, sums should be replaced by integrals.} $\cY$. The capacity of the channel, $I(W)$, is\footnote{The default basis of logarithms in this paper is 2} 
\begin{equation}
I(W) \defined \sum_{x\in\cX}\sum_{y\in\cY}0.5W\left(y\given x\right) \log \frac{W\left(y\given x\right)} {\sum_{x'\in\cX}0.5W\left(y\given x'\right)}\;.
\label{eq:Idef}
\end{equation}
Channel polarization \cite{arikan2009channel} is based on mapping two identical copies of the channel $W$ into the pair of BMS channels $W^-:\cX\rightarrow\cY^2$ and $W^+:\cX\rightarrow\cX\times\cY^2$ defined as
\begin{align}
W^-\left(y_1,y_2\given u_1\right)&=\sum_{u_2\in \cX}0.5W\left(y_1\given 
u_1\xor u_2\right)W\left(y_2\given u_1\right)\\W^+\left(y_1,y_2,u_1\given 
u_2\right)&=0.5W\left(y_1\given 
u_1\xor u_2\right)W\left(y_2\given u_2 \right)
\end{align}
Recalling that the channels $W^+$ and $W^-$ can be defined using density evolution operators,
$W^-=W\boxast W$ and $W^+=W\circledast W$
\cite{korada2009polar, mori2009per}, and applying~\cite[Theorem 4.141]{ru_book}, yields 
$
I^2(W)\le I\left(W^-\right) \le 1 - h_2\left\{ 2h_2^{-1}\left[1-I(W)\right]
\cdot\left[1-h_2^{-1}\left[1-I(W)\right]\right] \right\}
$
with $h_2(x)\defined-x\log x -(1-x)\log(1-x)$, and $h_2^{-1}$ is the inverse of $h_2$ with values in $\left[0,0.5\right]$.
In addition, $I(W^+)+I(W^-)=2I(W)$ \cite[Lemma 4.41]{ru_book}. This means $I\left(W^+\right)-I(W)=I(W)-I\left(W^-\right)\in\left[\epsilon_l\left(I(W)\right),\epsilon_h\left(I(W)\right) \right]$, where $\epsilon_h(x)=x-x^2$ and $\epsilon_l(x)=x-1+h_2\left\{h_2^{-1}\left(1-x\right)\left[2-2
h_2^{-1}\left(1-x\right)\right]\right\}$.
	
This procedure can now be reapplied to $W^-$ and $W^+$, creating $W^{--}$, $W^{-+}$, $W^{+-}$ and $W^{++}$. 
Repeating the procedure $\lambda$ times we obtain $2^\lambda$ BMS sub-channels, whose 
average capacity is $I(W)$. It was shown \cite{arikan2009channel} that these channels are polarized, i.e. for all $\delta\in(0,1)$
\begin{align}
\lim_{\lambda\rightarrow \infty}\left|\left\{
\bs\in\left\{+,-\right\}^n:I\left(W^\bs\right)\in(0,\delta)
\right\}\right|/2^\lambda
&=1-I(W)\\
\lim_{\lambda\rightarrow \infty}\left|\left\{
\bs\in\left\{+,-\right\}^n:I\left(W^\bs\right)\in(1-\delta,1)
\right\}\right|/2^\lambda
&=I(W)
\end{align}
Those $2^\lambda$ sub channels, denoted $W_0,W_1,\dots,W_{2^\lambda-1}$, are the  channels that the outer codes in the scheme described in Fig. \ref{fig:concat} see. Each outer code $A_i$ sees the channel $W_i$, and is designed to operate over this channel.

\section{Bounds on the error exponent} \label{sec:bounds_on_error_exps}
{Consider the concatenated, polarization based code that was described above.
The blocklength of the code is $N$ and it uses outer sub-codes with blocklength $N_1$. The number of polarization steps is $\lambda = \log (N / N_1)$.
We derive a lower bound on the error exponent (upper bound on the error rate) of the scheme by choosing the elements of the generating matrices of the linear sub-codes, $A_i$, uniformly at random, as described in \cite[Section 6.2]{galbook}.
We can obtain an upper bound on the average error probability of the ensemble of codes 
that we have just defined, $P_e$, using the successive decoding method outlined in Section \ref{sec:intro}, when transmitting over a 
given BMS channel, $W(y | x)$, as follows. We first compute the distributions (given that the zero codeword was transmitted) of the LLRs of the sub-channels after $\lambda$ polarization steps, using density evolution (DE) for polar codes as in 
\cite{korada2009polar,mori2009per}. Denote these distributions by $a_{W_i}(x)$, 
$i=0,1,\ldots,2^{\lambda}-1$. By~\cite[Sections 5.6 and
6.2]{galbook}, when using successive cancellation to decode the polar concatenated scheme described in Fig. \ref{fig:concat},
$$
P_e \le 
\sum_{i=0}^{2^{\lambda}-1} e^{-N_1 E_r(W_i,R_i)}
$$
where $W_i$ is the $i$th sub-channel, $R_i$ is the rate of the corresponding outer
sub-code, and $E_r(W,R)$ is the error exponent, given by\footnote{By \cite[Section 5.6]{galbook} $E_r(W,R) = \max_{0\le\rho\le 1} E_0(W,\rho) - \rho R$, where the rate $R$ is defined using natural logarithms, and measured in nats per channel use \cite[Paragraph after Equation (5.1.2)]{galbook}. In order to use the more widespread notation, by which $R$ is the number of bits per channel use, we substitute $R$ with $R\ln2$. In order to further improve the bound on $P_e$, we define $E_r(W,0)=\infty$ instead of the actual error exponent, since there cannot be errors if no information bits were transmitted.}
\begin{equation}
E_r(W,R) = \left\{
\begin{array}{ll}
\max_{0\le\rho\le 1} E_0(W,\rho) - \rho R{\cdot\ln 2} & R\ne 0 \\
\infty & R=0
\end{array}
\right. \label{eq:E_r_def}
\end{equation}
where
\begin{align}
\lefteqn{E_0(W,\rho) = -\ln \sum_y \left( 0.5
\left[W(y|0)^{\frac{1}{1+\rho}} + W(y|1)^{\frac{1}{1+\rho}} \right] \right)^{1+\rho}}\quad\: \: \\
&= (1+\rho)\ln 2 -\ln \sum_l 
\left[a_W(l)^{\frac{1}{1+\rho}} + a_W(-l)^{\frac{1}{1+\rho}} \right]^{1+\rho}
\end{align}
\comment{
\begin{align}
\lefteqn{E_0(W,\rho) = -\ln \sum_y \left( \frac{1}{2}\left[W(y\given 
		0)^{\frac{1}{1+\rho}} + \right. \right.}\\
&\qquad\qquad\qquad\qquad\qquad \left. \left. W(y\given 1)^{\frac{1}{1+\rho}} 
\right] \right)^{1+\rho}\\
&= (1+\rho)\ln 2 -\ln \sum_l 
\left[a_W(l)^{\frac{1}{1+\rho}} + a_W(-l)^{\frac{1}{1+\rho}} \right]^{1+\rho}
\end{align}
}
The second equality follows by modifying the original channel as in \cite[Lemma 4.35]{ru_book}: We add a 
processing block that computes the LLR from the original channel output. The 
new channel is also a BMS, and is operationally equivalent to the original channel.
	
According to \cite[Chapter 5.7]{galbook}, for low rates the average error probability is different from the typical error probability, since poor codes in the ensemble, although quite improbable, have a very high error probability. Using the expurgated error exponent provides a tighter estimate of the error probability of good codes than the random-coding exponent. This improved bound is \cite[Theorem 5.7.1]{galbook}. It asserts that the average error probability of the ensemble of typical codes with rate $R$ is upper bounded by $\exp\left[-NE_{ex}\left(W,R+\frac{2}{N}\right)\right]$, where
$$
E_{ex}(W,R)=\sup_{\rho\ge 1}E_x(W,\rho)-\rho R\cdot\ln 2
$$
and
\begin{align*}
E_x(W,\rho)&=-\rho\ln\left[\frac{1+\left(\sum_y\sqrt{W(y\given 0)W(y\given 1)}\right)^{\frac{1}{\rho}}}{2}\right]\\
&=-\rho\ln\left[1+\left(\sum_l\sqrt{a_W(l)a_W(-l)}\right)^{\frac{1}{\rho}}\right]+\rho\ln 2
\end{align*}
Therefore, in our polar concatenated scheme, we obtain the following upper bound on $P_e$, the ensemble average error probability when using only typical outer codes,
\begin{equation}
P_e \le 
\sum_{i=0}^{2^{\lambda}-1} e^{-N_1 E\left(W_i,R_i,N_1\right)}
\label{eq:err_exp_expur}
\end{equation}
where 
$$
E\left(W,R,N_1\right)=\max\left[E_r(W,R),E_{ex}(W,R+2/N_1)\right]
$$
This construction is called expurgated random code.
Define also 
\begin{equation}
E(W,R)\triangleq \lim_{N_1\rightarrow\infty}E\left(W,R,N_1\right)=\max\left[E_r(W,R),E_{ex}(W,R)\right] \label{eq:E(W,R)def}
\end{equation}
For the binary symmetric channel (BSC), the true error exponent of a typical linear code with rate $R$, transmitted over $W$, is $E(W,R)$ \cite{barg2002random}. It is conjectured to be the true exponent for other BMS channels as well.
By~\eqref{eq:err_exp_expur}, the error exponent, $E_\lambda(W,R)$, of the polarization-based code of blocklength $N$ with $\lambda$ polarization steps followed by $2^{\lambda}$ 
typical random linear codes of blocklength $N_1=N/2^{\lambda}$, with the best rate split, is lower bounded by, and conjectured to be equal to,
$$
\frac{1}{2^\lambda}{\max_{R_0,\dots,R_{2^{\lambda}-1}}}\min_{i,R_i\ne 0} E\left(W_i,R_i\right)
$$
where the maximization is over all possible combinations of rates 
$R_0,\dots,R_{2^{\lambda}-1}$ with total code rate $R$.
In~\cite{goldin2016block} we calculated this error exponent
by searching for those values of $R_i$ for which $E(W_i,R_i)$ are equal. We now present an improved approach that yields an explicit recursive expression for $E_\lambda(W,R)$ and produces the maximizing rates as a byproduct. Denote the minimal value of the right hand side (RHS) of \eqref{eq:err_exp_expur} by $\exp\left\{ -NE_\lambda(W,R,N_1) \right\}$ such that (s.t.)
\begin{equation}
E_\lambda(W,R,N_1) =\\ \frac{1}{N_1\cdot2^\lambda}\max_{R_0,\dots,R_{2^\lambda-1}}-\ln\sum_{i=0}^{2^\lambda-1} e^{-N_1 E\left(W_i,R_i,N_1\right)} \label{eq:E_p_def}
\end{equation}
where the maximization is over all possible combinations of rates, 
$R_0,\dots,R_{2^{\lambda}-1}$, s.t. $\sum_{i=0}^{2^\lambda-1} R_i = 2^{\lambda}\cdot R$, and for all $i$, $R_i\cdot N_1$ is an integer.
\begin{lemma} \label{lem:exact_recursion}
Define
$$
\mathcal{A}\triangleq \left\{x|x \in[\max(0,2R-1),\min(1,2R)], x\cdot N_1 2^{\lambda-1}\in\mathbb{Z}\right\}
$$
For any positive integers, $\lambda$, $N_1$ and $R N_1 2^\lambda$,
\begin{equation}
E_\lambda\left(W,R,N_1\right)=\max_{R_1\in\mathcal{A}}-\ln\left(e^{-N_12^{\lambda-1}E_{\lambda-1}\left(W^-,R_1,N_1\right)}+e^{-N_12^{\lambda-1}E_{\lambda-1}\left(W^+,2R-R_1,N_1\right)}\right)/\left(N_12^{\lambda}\right) \label{eq:exact_recursion}
\end{equation}
where $E_0(W,R,N_1)=E\left(W,R,N_1\right)$.
\end{lemma}
\begin{IEEEproof}
For $\lambda=1$ the claim follows immediately from \eqref{eq:E_p_def} for $\lambda=1$.
The condition $R_1\in[\max(0,2R-1),\min(1,2R)]$ follows from $R_1,2R-R_1\in[0,1]$.
		
For $\lambda>1$, note that $W_i$ are sub-channels of $W^-$ for $i=0,\dots,2^{\lambda-1}-1$, and sub-channels of $W^+$ for $i=2^{\lambda-1},\dots,2^\lambda-1$. \eqref{eq:E_p_def} can be rewritten as
\begin{align}
E_\lambda\left(W,R,N_1\right) 
&= 
\frac{1}{N_1 2^{\lambda}}\max_{R'\in\cA}
\max_{\substack{R_0\dots R_{2^{\lambda}-1}\\
\sum_{i=0}^{2^{\lambda-1}-1} R_i=R'\cdot 2^{\lambda-1}\\ \sum_{i=2^{\lambda-1}}^{2^\lambda-1} R_i=(2R-R')\cdot 2^{\lambda-1}}}
-\ln\sum_{i=0}^{2^\lambda-1} e^{-N_1 E\left(W_i,R_i,N_1\right)}\\
&= 
\frac{1}{N_1 2^{\lambda}}
\max_{R'\in\cA}-\ln\left(
\min_{\substack{R_0,\dots,R_{2^{\lambda-1}-1}\\ \sum R_i=R'\cdot 2^{\lambda-1}}}
\sum_{i=0}^{2^{\lambda-1}-1} e^{-N_1 E\left(W_i,R_i,N_1\right)} +
\min_{\substack{R_{2^{\lambda-1}},\dots,R_{2^\lambda-1}\\\sum R_i=(2R-R')\cdot 2^{\lambda-1}}} \sum_{i=2^{\lambda-1}}^{2^\lambda-1}e^{-N_1E\left(W_i,R_i,N_1\right)}\right)\\
&=
\frac{1}{N_12^{\lambda}}\max_{R'\in\cA}-\ln\left(e^{-N_1\cdot2^{\lambda-1}E_{\lambda-1}\left(W^-,R',N_1\right)}+e^{-N_1\cdot2^{\lambda-1}E_{\lambda-1}\left(W^+,2R-R',N_1\right)}\right)
\end{align}
where the first equality follows from rewriting \eqref{eq:E_p_def}, the second one follows from splitting the inner maximization into two separate ones and inserting them into the monotonic decreasing function $-\ln()$, and the third equality follows from applying \eqref{eq:E_p_def} for $\lambda-1$ instead of $\lambda$.
\end{IEEEproof}
We conjecture that the condition $R_1\in[\max(0,2R-1),\min(1,2R)]$ in \eqref{eq:exact_recursion} can be replaced by $R_1\in[\max(0,2R-1),R]$. This is due to the fact that, compared to $W^-$, $W^+$ is a better channel. Hence the information rate of the sub-code corresponding to $W^+$ should be larger than the rate of the sub-code corresponding to $W^-$.

Denote
\begin{equation}
\cB\triangleq [\max(0,2R-1),\min(1,2R)]
\label{eq:Bdef}
\end{equation}
\begin{lemma}
Define
$E_0(W,R)=E(W,R)$ (defined in \eqref{eq:E(W,R)def}), and for $\lambda\ge 1$ define recursively,
\begin{equation}
E_\lambda(W,R)=0.5\max_{R_1\in\cB}E_{m,\lambda}\left(W,R,R_1\right)
\label{eq:maxmin_recursion}
\end{equation} 
where
\begin{equation}
\label{eq:Eml_WRR1}
E_{m,\lambda}\left(W,R,R_1\right)\triangleq \min[E_{\lambda-1}\left(W^-,R_1\right),E_{\lambda-1}\left(W^+,2R-R_1\right)]
\end{equation}
Then for $\lambda\ge 0$,
$$
E_\lambda\left(W,R\right)-\Theta\left(\frac{1}{N_1}\right)\le E_\lambda\left(W,R,N_1\right)\le E_\lambda(W,R)
$$
\label{lem:e_lamWRN1}
\end{lemma}
The bound shows that for large $N_1$, $E_\lambda(W,R,N_1)$ approaches $E_\lambda(W,R)$.
\begin{IEEEproof}
We prove by induction that $E_\lambda\left(W,R,N_1\right)\le E_\lambda\left(W,R\right)$, with $E_\lambda\left(W,R\right)$ defined recursively in \eqref{eq:maxmin_recursion}.
This claim is true for $\lambda=0$ since
\begin{align}
E_0\left(W,R,N_1\right) &= E\left(W,R,N_1\right) =\max \left[E_r(W,R),E_{ex}\left(W,R+2/N_1\right)\right]\\
&\le \max \left[E_r(W,R),E_{ex}\left(W,R\right)\right]=E(W,R)=E_0(W,R)
\end{align}
where the inequality is due to the fact that $E_{ex}(W,R)$ is a decreasing function of $R$.
For shortness of notation, define 
$$
E_{m,\lambda}\left(W,R,R_1,N_1\right)\triangleq \min\left[E_{\lambda-1}\left(W^-,R_1,N_1\right),E_{\lambda-1}\left(W^+,2R-R_1,N_1\right)\right]
$$
Assuming the claim is true for $\lambda-1$, we have
\begin{align}
E_\lambda\left(W,R,N_1\right)
&\le \frac{1}{N_12^\lambda}\max_{R_1\in\cA}-\ln\left(e^{-N_1\cdot 2^{\lambda-1}E_{m,\lambda}\left(W,R,R_1,N_1\right)]}\right)\\
&= 0.5\max_{R_1\in\cA}E_{m,\lambda}\left(W,R,R_1,N_1\right)
\le 0.5\max_{R_1\in\cA}E_{m,\lambda}\left(W,R,R_1\right)\\
&\le 0.5\max_{R_1\in\cB}E_{m,\lambda}\left(W,R,R_1\right)=E_\lambda(W,R)
\end{align}
where the first inequality follows from \eqref{eq:exact_recursion}, and the induction assumption yields the second inequality.
		
Similarly, we can prove that $E_\lambda\left(W,R\right)-\Theta\left(\frac{1}{N_1}\right)\le E_\lambda\left(W,R,N_1\right)$ using induction:
The claim is true for $\lambda=0$ since 
$$
E_{ex}\left(W,R+2/N_1\right)\ge E_{ex}\left(W,R\right)-\frac{2}{N_1}\left|\frac{\partial E_{ex}(W,r)}{\partial r}\right|_{r=R}=E_{ex}\left(W,R\right)-\Theta\left(\frac{1}{N_1}\right)
$$
where the inequality follows from the fact that $E_{ex}(W,R)$ is a convex and decreasing function of $R$. This yields
\begin{align}
E_0\left(W,R,N_1\right) &= \max \left[E_r(W,R),E_{ex}\left(W,R+2/N_1\right)\right]\\
&\ge\max \left[E_r(W,R),E_{ex}\left(W,R\right)\right]-\Theta\left(\frac{1}{N_1}\right)=E_0(W,R) -\Theta\left(\frac{1}{N_1}\right)
\end{align}
Assuming that the claim is true for $\lambda-1$, we have
\begin{align}
E_\lambda&\left(W,R,N_1\right)\\&\ge \frac{1}{N_12^\lambda}\max_{R_1\in\cA}-\ln\left(2e^{-N_1\cdot 2^{\lambda-1}E_{m,\lambda}\left(W,R,R_1,N_1\right)}\right)\\
&=0.5\max_{R_1\in\cA}E_{m,\lambda}\left(W,R,R_1,N_1\right)-{\ln 2}/\left(N_12^\lambda\right)\\
&\ge 0.5\max_{R_1\in\cA}E_{m,\lambda}\left(W,R,R_1\right)-\Theta\left({1}/{N_1}\right) \label{eq:pre_last_max}
\end{align}
where the first inequality follows from \eqref{eq:exact_recursion}, and the second follows from the induction assumption.
		
Define
$$
\hat{R_1}\triangleq \argmax_{R_1\in \cB}
E_{m,\lambda}(W,R,R_1)
$$
Since $E_{\lambda-1}\left(W^-,R_1\right)$ and $E_{\lambda-1}\left(W^+,2R-R_1\right)$ are convex functions of $R_1$ (see Lemma \ref{lem:f_property} below, Appendix~\ref{app:lem_f_property_proof} and Fig. \ref{fig:required}),
\begin{align}
E_{m,\lambda}(W,R,R_1) &= \min\left[E_{\lambda-1}\left(W^-,R_1\right),E_{\lambda-1}\left(W^+,2R-R_1\right)\right]\\
&\ge \left\{\begin{array}{cc}
E_{\lambda-1}\left(W^-,\hat{R_1}\right)-\left|\frac{\partial E_{\lambda-1}\left(W^-,r\right)}{\partial r}\right|_{r=\hat{R_1}}(R_1-\hat{R_1})& R_1>\hat{R_1}\\
E_{\lambda-1}\left(W^+,2R-\hat{R_1}\right)+\left|\frac{\partial E_{\lambda-1}\left(W^+,2R-r\right)}{\partial r}\right|_{r=\hat{R_1}}(R_1-\hat{R_1})& R_1<\hat{R_1}
\end{array}
\right.\\
&= 2E_{\lambda}\left(W,R\right)-\left\{\begin{array}{cc}
\left|\frac{\partial E_{\lambda-1}\left(W^-,r\right)}{\partial r}\right|_{r=\hat{R_1}}(R_1-\hat{R_1})& R_1>\hat{R_1}\\
\left|\frac{\partial E_{\lambda-1}\left(W^+,2R-r\right)}{\partial r}\right|_{r=\hat{R_1}}(\hat{R_1}-R_1)& 
R_1<\hat{R_1}
\end{array}
\right.
\end{align}
Maximizing this expression over $R_1\in[\max(0,2R-1),\min(1,2R)]$ s.t. $R_1\cdot N_1 2^{\lambda-1}\in\mathbb{Z}$ yields a result for $R_1-\hat{R_1}=\Theta\left(\frac{1}{N_1 2^{\lambda-1}}\right)=\Theta\left(\frac{1}{N_1}\right)$. Therefore, the result of the maximization is $2E_{\lambda}\left(W,R\right)-\Theta\left(\frac{1}{N_1}\right)$. Combining this with \eqref{eq:pre_last_max} yields $E_{\lambda}\left(W,R,N_1\right)\ge E_{\lambda}\left(W,R\right)-\Theta\left(\frac{1}{N_1}\right)$.
\end{IEEEproof}
\begin{lemma} \label{lem:f_property}
For any integer $\lambda\ge 0$, and any BMS $W$, $E_{\lambda}(W,R)$ is a finite, decreasing and convex function of $R$ for $R>0$. Furthermore, for $R>0$, $\left|\partial E_{\lambda}(W,R)/\partial R\right|<\infty.$
\end{lemma}
We prove the lemma in Appendix \ref{app:lem_f_property_proof}.

In Fig. \ref{fig:exponents} we plot the resulting error exponent $E_\lambda(W,R)$ for a code with rate $R=1/2$ and $\lambda=1,2,3$ as a function of the SNR
when transmitting over a BIAWGNC.
\begin{figure} 
\centering
\includegraphics[width=0.5\columnwidth]{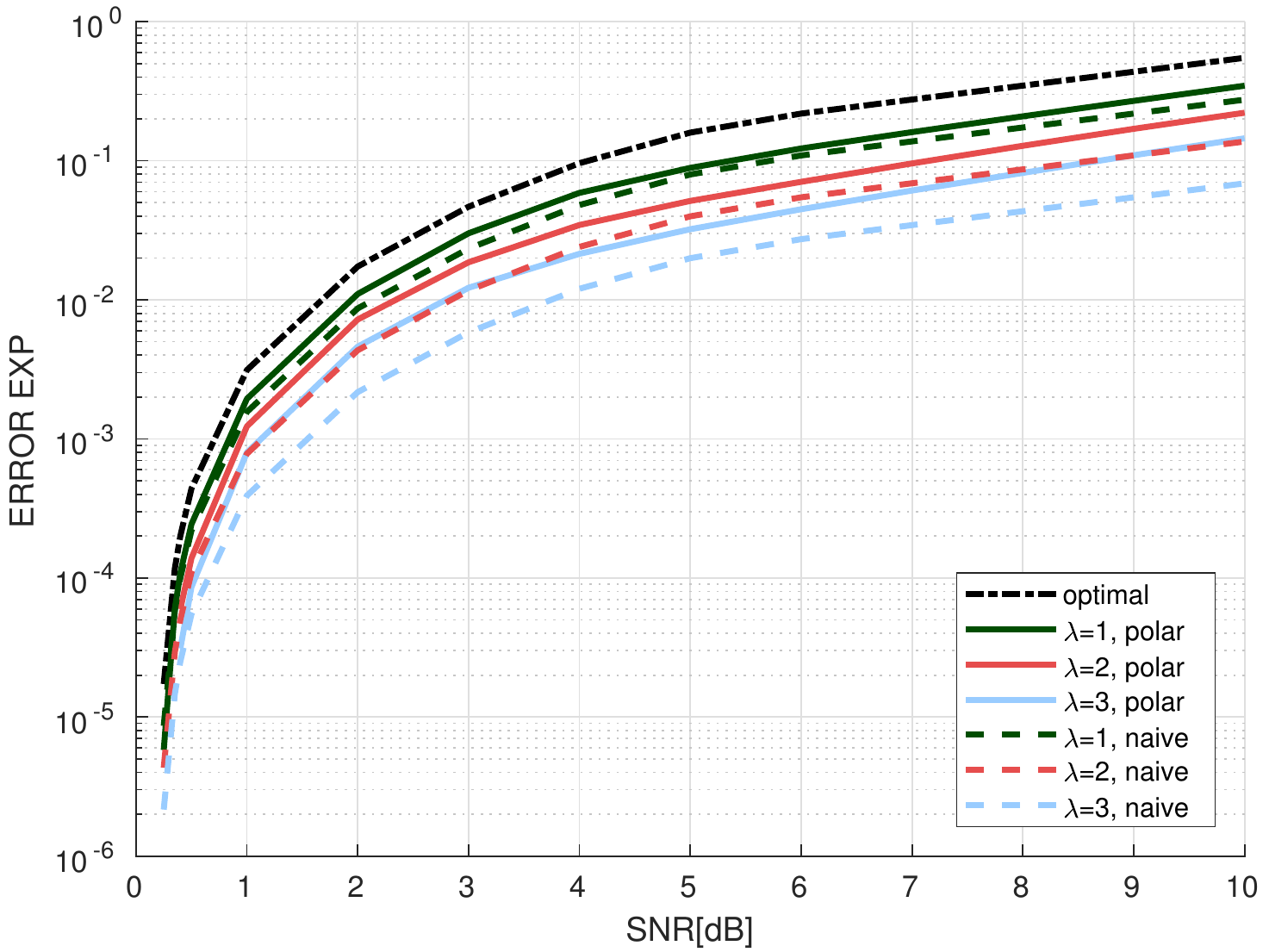} 
\vspace{-10pt}
\caption{Error exponents of codes created by $\lambda$ polar transforms 
followed by $2^{\lambda}$ codewords of outer codes with blocklength $N/2^{\lambda}$ and 
optimal rate-splitting, and naively using $2^{\lambda}$ codewords of a code with
blocklength $N/2^{\lambda}$. The channel is BIAWGNC, and the total 
rate is $1/2$.}
\label{fig:exponents}
\end{figure}
This is compared to a naive approach where we simply use $2^{\lambda}$ 
codewords of a code with blocklength $N_1=N/2^{\lambda}$ (without using the polar 
transformation at all).
Denote the error exponent of the polar-based (naive, respectively) approach by 
$E_\lambda$ and $E_{n,\lambda}$.
We also plot the error exponent of an optimal code, $E$, corresponding to 
$\lambda=0$.
Note that $E_{n,\lambda}=E/2^{\lambda}$. As expected, polarization is useful, i.e. $E_\lambda\ge E_{n,\lambda}$. All the plots in Fig. \ref{fig:exponents} have an asymptote at the SNR corresponding to the capacity $C=R=1/2$, which equals 0.19dB, since for this SNR, $E$, $E_\lambda$ and $E_{n,\lambda}$ approach zero.
This explains why in all the discussed codes, the gap to the optimal error 
exponent plot is small for low SNR.

Although as the channel capacity increases (as in the case of high SNR), polarization is less effective (because for high capacity, the polarization gap, $I(W^+)-I(W)=I(W)-I(W^-)$, approaches zero, and therefore polarization has almost no affect on the channel), $E_{n,\lambda}/E_\lambda\nrightarrow 1$ for high SNR. This is due to the fact, that for expurgated ensembles $\lim_{SNR\rightarrow\infty}E(W,R)=\infty$, so $E_{n,\lambda}$ and $E_\lambda$ approach infinity for high SNR. It should be noted, that for the random coding ensemble $\lim_{SNR\rightarrow\infty}E_r(W,R)=\ln 2 (1-R)$, so if we have not considered expurgated ensembles (i.e., ensembles with typical outer codes only), we would have obtained $E_{n,\lambda}/E_\lambda\rightarrow 1$ for high SNR.
It should also be noted that the error exponent improvement increases with 
$\lambda$, i.e, for a desired error exponent, the SNR improvement by applying 
$\lambda$ polarization steps, compared to the naive approach, increases with 
$\lambda$.
For example, for desired error exponent of $0.01$, the SNR gain for 
$\lambda=1,2,3$ is $0.2{\rm dB}, 0.5{\rm dB}$ and $0.95{\rm dB}$ respectively.
	

Fig. \ref{fig:exponents} demonstrates that polarization improves the error exponent compared to the naive approach. In the following we provide some partial theoretical justification. 
Define $E_{r,\lambda}(W,R)$ the same way as $E_\lambda(W,R)$ in Lemma \ref{lem:e_lamWRN1}, except that $E_{r,0}(W,R)=E_r(W,R)$. That is, we only consider a random coding (non-expurgated) error exponent analysis. Similarly, define $E_{n,r,\lambda}(W,R)\triangleq E_r(W,R)/2^\lambda$. Using the results in \cite{alsan2014polarization}, we claim that under certain conditions, $E_{r,\lambda}(W,R)\ge E_{n,r,\lambda}(W,R)$.
We demonstrate this for the case $\lambda=1$, although the argument may be extended to larger values of $\lambda$.
The combination of
\begin{equation}
2E_0(W,\rho)\le E_0\left(W^+,\rho\right)+E_0\left(W^-,\rho\right) \label{eq:alsan}
\end{equation} 
\cite{alsan2014polarization} and \eqref{eq:E_r_def} yields that for $R\ne 0$ and all $R_1$:
\begin{align}
2E_r(W,R)
&=
\max_{\rho\in[0,1]} 2\left[E_0(W,\rho) - \rho R \ln 2 \right] \le \max_{\rho\in[0,1]} \left[E_0(W^+,\rho) - \rho R_1 \ln 2 + E_0\left(W^-,\rho\right) - \rho\left(2R-R_1\right) \ln 2 \right] 
\label{eq:recursion_alsan}\\
&\le
\max_{\rho\in[0,1]} \left[E_0\left(W^+,\rho\right) - \rho R_1 \ln 2 \right] + 
\max_{\rho\in[0,1]} \left[E_0\left(W^-,\rho\right) - \rho \left(2R-R_1\right) \ln 2 \right]\\
&=
E_r\left(W^-,R_1\right)+E_r\left(W^+,2R-R_1\right)
\end{align}
This is true in particular for $R_1=\hR_1$, defined by
$$
\hat{R}_1\triangleq\argmax_{R_1 \in \cB}
\min\left[E_r\left(W^-,R_1\right),E_r\left(W^+,2R-R_1\right)\right]
$$
Now, $E_r\left(W^-,R_1\right)$ is decreasing in $R_1$, and $E_r\left(W^+,2R-R_1\right)$ is increasing in $R_1$.
If the following condition (which parallels \eqref{eq:mid_rates} in Appendix \ref{app:lem_f_property_proof})
\begin{equation}
\left\{\begin{array}{l}
E_{r}\left(W^-,0^+\right)\ge E_{r}\left(W^+,2R\right)\\
E_{r}\left(W^-,2R\right)\le E_{r}\left(W^+,0^+\right)
\end{array}\right. \label{eq:mid_rates1}
\end{equation}
holds (where for any $W$ and $R>1$ we define $E_r(W,R)\equiv 0$), then $E_r\left(W^-,R_1\right)$ and $E_r\left(W^+,2R-R_1\right)$ intersect at $\hR_1$ (see Fig. \ref{fig:required} in Appendix \ref{app:lem_f_property_proof} with $E_r(W^-,r)$ replaced by $E_{\lambda-1}(W^-,r)$ and $E_r(W^+,2R-r)$ replaced by $E_{\lambda-1}(W^+,2R-r)$). In this case 
$$
E_r\left(W^-,\hat{R}_1\right)+E_r\left(W^+,2R-\hat{R}_1\right)=2\max_{R_1}\min\left[E_r\left(W^-,R_1\right),E_r\left(W^+,2R-R_1\right)\right] = 4E_{r,1}(W,R)
$$
But the LHS of \eqref{eq:recursion_alsan} equals $4E_{n,r,1}(W,R)$.
We conclude that if \eqref{eq:mid_rates1} holds then $E_{r,1}(W,R)\ge E_{n,r,1}(W,R)$. However, if \eqref{eq:mid_rates1} does not hold then either $\hR_1 = \max(0,2R-1)$ or $\hR_1 = \min(1,2R)$ and the argument fails.

We now obtain a lower bound on the achievable error exponent $E_\lambda(W,R)$ that depends on the channel, $W$, only through its capacity, $I(W)$.
By \cite[Equation (26)]{guillen2013extremes}, for a given channel capacity $I(W)$,
\begin{equation}
E_{r,BSC}\left(I(W),R\right)\le E_r\left(W,R\right)\le E_{r,BEC}\left(I(W),R\right)
\end{equation}
where $E_{r,BEC}\left(I(W),R\right)$ ($E_{r,BSC}\left(I(W),R\right)$, respectively) is the error exponent corresponding to random codes of rate $R$ over a binary erasure channel (binary symmetric channel) of capacity $I(W)$. By \cite[Equation (27), Theorem 2]{guillen2013extremes} this is also true for expurgated error exponents. Therefore,
\begin{equation}
E_{BSC}\left(I(W),R\right)\le E\left(W,R\right)\le E_{BEC}\left(I(W),R\right)
\end{equation}
is also true for expurgated random codes and their error exponents as defined in \eqref{eq:E(W,R)def}.
Recalling that
\begin{equation}
I\left(W^+\right)-I(W)=I(W)-I\left(W^-\right)\in\left[\epsilon_l\left(I(W)\right),\epsilon_h\left(I(W)\right) \right]
\end{equation}
(see Section \ref{sec:background}), we define
\begin{equation}
\hat{E}_{\lambda}\left(I(W),R\right)\triangleq 0.5\min_{\epsilon\in\left[\epsilon_l,\epsilon_h\right]}\max_{R_1\in\cB} \hat{E}_{m,\lambda}\left(I(W),\epsilon,R,R_1\right)
\end{equation}
where $\epsilon_l=\epsilon_l(I(W))$, $\epsilon_h=\epsilon_h(I(W))$,
$$
\hat{E}_{m,\lambda}\left(I(W),\epsilon,R,R_1\right)=\min\left[\hat{E}_{\lambda-1}\left(I(W)-\epsilon ,R_1\right),\hat{E}_{\lambda-1}\left(I(W)+\epsilon ,2R-R_1\right)\right]
$$ and $\hat{E}_{0}\left(I(W),R\right)=E_{BSC}\left(I(W),R\right)$ Note that $\hat{E}_{\lambda}\left(I(W),R\right)$ does not depend on the channel $W$, but only on its capacity.
\begin{theorem}
For any BMS channel $W$ with capacity $I(W)$, any desired code rate $R$, and a concatenated code with $\lambda\ge 0$ polarization steps for the inner polar code and (randomly generated linear) outer codes with blocklength $N_1\rightarrow\infty$, the best achievable error exponent $E_\lambda\left(W,R\right)$ is lower bounded by 
$E_\lambda\left(W,R\right)\ge\hat{E}_{\lambda}\left(I(W),R\right)$. 
\end{theorem}
\begin{IEEEproof}
We will prove the theorem using induction. The claim is trivial for $\lambda=0$. We will prove it for $\lambda$ assuming it is true for $\lambda -1$.
\begin{align*}
E_\lambda(W,R) 
&= 0.5\max_{R_1\in\cB}E_{m,\lambda}\left(W,R,R_1\right)\\
&\ge 0.5\max_{R_1\in\cB}\min\left[\hat{E}_{\lambda-1}\left(I\left(W^-\right),R_1\right),
\hat{E}_{\lambda-1}\left(I\left(W^+\right),2R-R_1\right)\right]\\
&\ge 0.5\min_{\epsilon\in\left[\epsilon_l,\epsilon_h\right]}\max_{R_1\in\cB}\hat{E}_{m,\lambda}\left(I(W),\epsilon,R,R_1\right)
=\hat{E}_{\lambda}\left(I(W),R\right)
\end{align*}
where the first equality follows from \eqref{eq:maxmin_recursion}, the first inequality follows from the induction assumption, and the second one follows from the fact that there exists $\epsilon_l\left(I(W)\right)\le\epsilon\le \epsilon_h\left(I(W)\right)$ s.t. $I(W)-I\left(W^-\right)=I\left(W^+\right)-I(W)=\epsilon$. 
\end{IEEEproof}
We thus showed that $\hat{E}_{\lambda}\left(I(W),R\right)$ is a lower bound on the error exponent $E_\lambda(W,R)$ for all BMS channels with a given capacity $I(W)$.
In Fig. \ref{fig:exponents3I}, we plot the asymptotic lower bound for the error exponent, $\hat{E}_{\lambda}\left(I(W),R\right)$, for a code with rate $R=1/2$ and $\lambda=1,2,3$ as a function of $I(W)$. This is compared to the lower bound for the naive approach, where we use $2^{\lambda}$ codewords with blocklength $N_1=N/2^{\lambda}$, without using the polar transformation. This lower bound is $E_{BSC}\left(I(W),R\right)/2^{\lambda}$. As expected, the lower bound for the scheme with polarization is higher, i.e. $\hat{E}_{\lambda}\ge \hat{E}_{n,\lambda}$. All the plots in Fig. 
\ref{fig:exponents3I} have an asymptote at 
$I(W)=R=1/2$, since when the rate approaches capacity, $E_{BSC}$, $\hat{E}_{\lambda}$ and $\hat{E}_{n,\lambda}$ 
approach zero.
We also see that $\hat{E}_{\lambda}$ and $\hat{E}_{n,\lambda}$ have an asymptote at $I(W)=1$. This asymptote follows from using expurgated codes, and since it is a lower bound on the error exponent, it follows that $E_\lambda$ and $E_{n,\lambda}$ approach infinity as $\rm{SNR}\rightarrow\infty$ ($I(W)\rightarrow 1$) in Fig. \ref{fig:exponents} for the BIAWGNC.
It should also be noted that the difference between the lower bounds on the error exponents of the polarization-based and naive schemes increases with 
$\lambda$.
\begin{figure} 
\centering
\includegraphics[width=0.5\columnwidth]{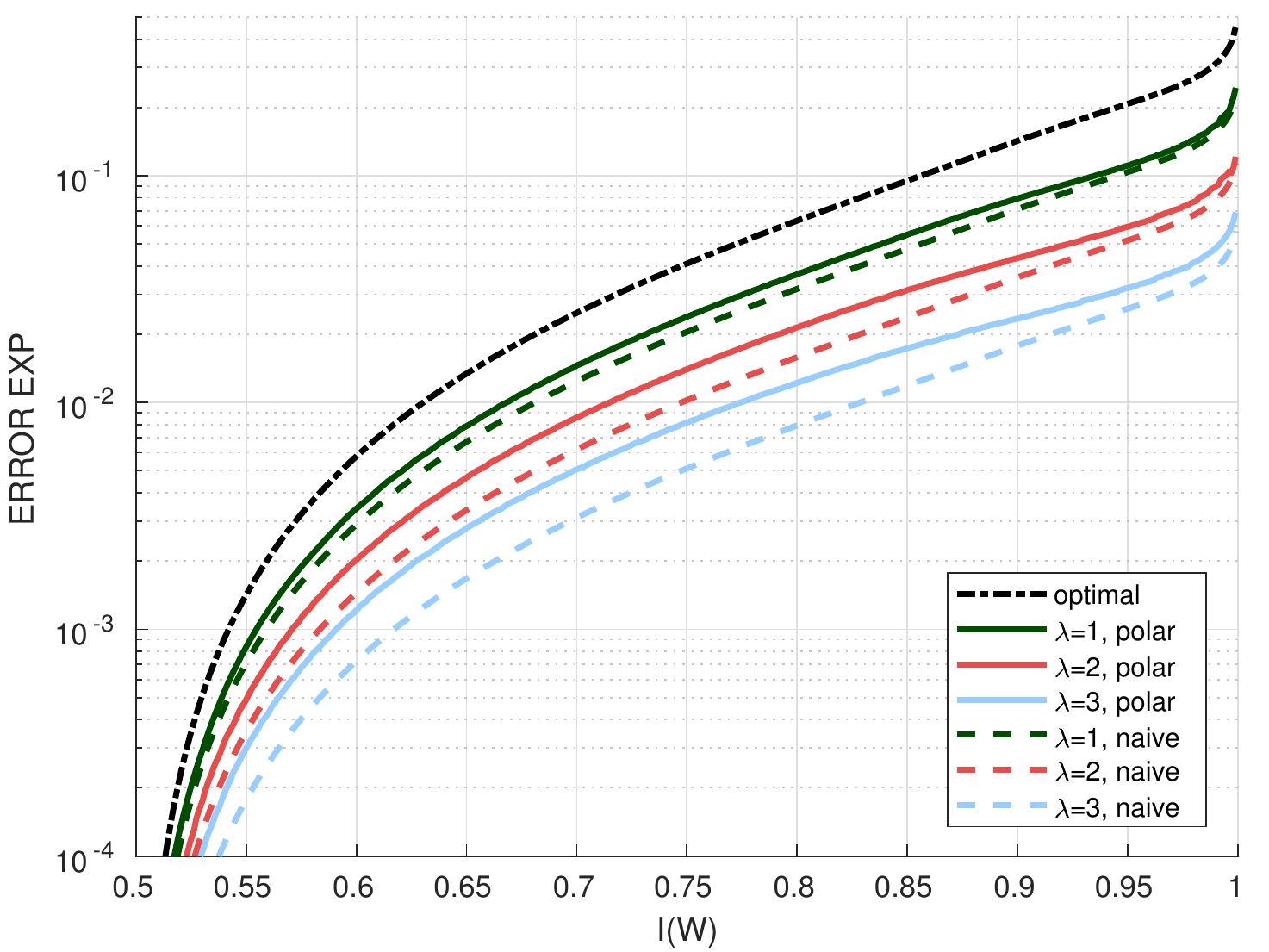} 
\vspace{-10pt}
\caption{Lower bounds for error exponents of codes created by $\lambda$ polar transforms 
followed by $2^{\lambda}$ outer codes of blocklength $N/2^{\lambda}$ and rate $R=1/2$ with optimal rate-splitting, and naively using $2^{\lambda}$ outer codes of 
blocklength $N/2^{\lambda}$, for a general BMS channel with capacity $I(W)$ and total rate $R=1/2$.} \label{fig:exponents3I}
\end{figure}

\section{Improved bounds and approximations} \label{sec:improved_bnds_and_approx}
\subsection{Achievable bound for BEC} \label{sec:improved_achievable_BEC}
\cite[Theorem 37]{polyanskiy2010channel} provides an upper bound on the achievable frame error rate (FER) of an error-correcting code with blocklength $N$ and rate $R$ for the binary erasure channel (BEC) with erasure probability $\epsilon$.
Combining this upper bound with  \eqref{eq:err_exp_expur}, we obtain that for a polar-concatenated code with outer codes of length $N_1$, and rates $R_i$, $i=0,\ldots,2^\lambda -1$, an achievable upper bound is
\begin{align}
P_e 
&\le
\sum_{i=0}^{2^\lambda-1}\min\left\{\sum_{t=0}^{N_1} \binom{N_1}{t} (1-I(W_i))^t I(W_i)^{N_1-t}
2^{-\left[N_1-t-\log\left(\frac{2^{N_1R_i}-1}{2}\right)\right]^+},e^{-N_1E\left(W_i,R_i,N_1\right)}\right\} \label{eq:theo37}\\
&\defined
\sum_{i=0}^{2^\lambda-1} \gamma_i(R_i)
\end{align}
where $x^+\triangleq \max(x,0)$.
We compute an upper bound on the achievable FER for polar codes of length $2^\lambda$ concatenated with codes of length $N_1$ and the best rate division between sub-codes given the total code rate $R$, by minimizing the RHS of \eqref{eq:theo37} given $\sum_{i=0}^{2^\lambda-1} R_i=2^\lambda R$ (which can also be written as $\sum_{i=0}^{2^\lambda-1} R_i N_1 = R N$) using the following efficient dynamic programming algorithm. Denote
$$
\delta_l(\rho) \defined \min_{R_0,\ldots,R_{l}} \sum_{i=0}^{l} \gamma_i(R_i)
$$
for integer $\rho\in [0,R N]$. The minimization is under the constraint that $N_1 R_i$ are all positive integers, satisfying $\sum_{i=0}^{l} R_i N_1 = \rho$. 
For each $l=0,1,\ldots,2^\lambda-1$, and each integer value of $\rho\in [0,R N]$, the algorithm computes $\delta_l(\rho)$ recursively using
$$
\delta_{l}(\rho) = \min_{R_l} \left\{ \delta_{l-1}(\rho - N_1 R_l) + \gamma_l(R_l) \right\}
$$
subject to the constraint that $N_1 R_l$ is an integer and $N_1 R_l \in [0,\rho]$. The recursion is initialized using
$$
\delta_0(\rho) = \gamma_0(\rho/N_1)
$$
The output of the algorithm is $\delta_{2^\lambda-1}(RN)$, which is the minimum of the RHS of \eqref{eq:theo37}. The minimizing rates, $R_0,\ldots,R_{2^\lambda-1}$, can be easily obtained as a byproduct of this recursive algorithm.
\subsection{Dispersion-based (normal) approximation} \label{sec:dispersion}
Consider transmission over a BMS channel, $W$, with capacity $I(W)$ and dispersion $V(W)$, defined as $V(W) \defined \sum_{x\in\cX}\sum_{y\in\cY}\frac{1}{2}W\left(y\given x\right) \left( \log \frac{W\left(y\given x\right)} {\sum_{x'\in\cX}\frac{1}{2}W\left(y\given x'\right)} \right)^2 - I(W)^2$.
By~\cite{polyanskiy2010channel}, the maximal rate of transmission at error probability $\epsilon$ and blocklength $N$ is closely approximated by $I(W)-\sqrt{\frac{V(W)}{N}}Q^{-1}(\epsilon)$ where $Q()$ is the complementary Gaussian cumulative distribution function. The approximation improves as $N$ gets larger, but is known to be tight already for $N$ as short as about 100.
The error probability of the best code with blocklength $N$ and rate $R$ can be approximated by $Q\left\{\sqrt{\frac{N}{V(W)}}\left[I(W)-R+O\left(\frac{\log N}{N}\right)\right]\right\}$ \cite{polyanskiy2010channel} (this is sometimes referred to as the {\em normal approximation} in the literature). However, for some channels this expression can be improved. For BIAWGNC \cite{erseghe2016coding} and BSC \cite[Theorem 52]{polyanskiy2010channel}, $P_e\approx Q\left\{\sqrt{\frac{N}{V(W)}}\left[I(W)-R+\frac{\log N}{2N}+O\left(N^{-1}\right)\right]\right\}$, and for BEC it is the same expression without $0.5\log N/N$ \cite[Theorem 53]{polyanskiy2010channel}. Therefore, for the BEC and for general channels we approximate the error probability as $Q\left(\sqrt{\frac{N}{V(W)}}\left(I(W)-R\right)\right)$, while for the BIAWGNC and BSC we use 
\begin{equation}
Q\left[\sqrt{\frac{N}{V(W)}}\left(I(W)-R+\frac{\log N}{2N}\right)\right] \label{eq:Q_biawgn}
\end{equation}
The smallest achievable error probability of our concatenated polar coding scheme is thus approximated by
\begin{align}
P_e\approx\min_{\substack{R_0\dots R_{2^{\lambda}-1}\\R_i\in\left[0,I\left(W_i\right)\right]\\\sum_{i=0}^{2^\lambda-1}R_i=2^\lambda R}}\sum_{i=0}^{2^\lambda-1} Q\left(\left(I\left(W_i\right)-R_i+C(N_1)\right)\sqrt{\frac{N_1}{V(W_i)}}\right)\label{eq:V_approx}\\
\le 2^\lambda \min_{\substack{R_0\dots R_{2^{\lambda}-1}\\R_i\in\left[0,I\left(W_i\right)\right]\\\sum_{i=0}^{2^\lambda-1}R_i=2^\lambda R}}\max_i Q\left(\left(I\left(W_i\right)-R_i+C(N_1)\right)\sqrt{\frac{N_1}{V(W_i)}}\right) \label{eq:V_upper}
\end{align}
where $C(N_1)$ is a correction term. For the BIAWGNC and BSC, $C(N_1)=\log N_1 / (2N_1)$, and for other channels $C(N_1)=0$.
\rem{
\begin{equation}
P_e\approx\min_{\substack{R_0\dots R_{2^{\lambda}-1}\\R_i\in\left[0,I\left(W_i\right)\right]\\\sum_{i=0}^{2^\lambda-1}R_i=2^\lambda R}}\sum_{i=0}^{2^\lambda-1} Q\left(\left(I\left(W_i\right)-R_i+\frac{\log N_1}{2N_1}\right)\sqrt{\frac{N_1}{V(W_i)}}\right)\label{eq:V_approx_awgn}
\end{equation}
}
Note that $P_e$ is also approximately lower bounded by the same expression in \eqref{eq:V_upper} without the multiplying $2^\lambda$ term.

In the simulations section it will be shown that \eqref{eq:V_approx} provides a tight approximation to the actual performance of concatenated BCH-polar codes over the BIAWGNC. The minimization in \eqref{eq:V_approx} is computed efficiently using a dynamic programming algorithm, similarly to the algorithm that was described in Section \ref{sec:improved_achievable_BEC}. The algorithm also provides the rates $R_0,\ldots,R_{2^\lambda-1}$ as a byproduct.

To gain insight on the min-max problem \eqref{eq:V_upper}, we first look at the simpler problem
\begin{equation}
\min_{\substack{R_0\dots R_{2^{\lambda}-1}\\\sum_{i=0}^{2^\lambda-1}R_i=2^\lambda R}}\max_i Q\left(\left(I\left(W_i\right)-R_i\right)\sqrt{\frac{N_1}{V(W_i)}}\right) \label{eq:simplified_V}
\end{equation}
This problem is solved when $\frac{I\left(W_i\right)-R_i}{\sqrt{V(W_i)}}$, $i=0,\dots,2^\lambda-1$ are equal and $\sum_{i=0}^{2^\lambda-1}R_i=2^\lambda R$. 
Hence, $I\left(W_i\right)-R_i=\frac{2^\lambda[I(W)-R]\sqrt{V\left(W_i\right)}}{\sum_{j=0}^{2^\lambda-1}\sqrt{V\left(W_j\right)}}$, and the solution of \eqref{eq:simplified_V} is 
$Q\left([I(W)-R]\sqrt{\frac{N}{V_{\lambda}(W)}}\right)$ where 
\begin{equation}
V_{\lambda}(W) \defined \frac{\left(\sum_{i=0}^{2^\lambda-1}\sqrt{V\left(W_i\right)}\right)^2}{2^\lambda} 
\label{eq:Vp_def}
\end{equation}
Since \eqref{eq:simplified_V} is a relaxed version of the min-max problem in \eqref{eq:V_upper}, if the solution of \eqref{eq:simplified_V} obeys all the constraints of the min-max problem in \eqref{eq:V_upper}, then it is the solution of this problem as well. Therefore, if $\forall i=0,\dots,2^\lambda-1$
\begin{equation}
R_i=I\left(W_i\right)-\frac{2^\lambda[I(W)-R]\sqrt{V\left(W_i\right)}}{\sum_{j=0}^{2^\lambda-1}\sqrt{V\left(W_j\right)}}\ge 0 \label{eq:condition}
\end{equation}
then
\begin{equation}
P_e \approx
2^\lambda Q\left([I(W)-R]\sqrt{\frac{N}{V_{\lambda}(W)}}\right)
\label{eq:Pe_approx_polar}
\end{equation}
For $R$ sufficiently close to $I(W)$ the condition \eqref{eq:condition} is guaranteed to hold. The minimal $R$, for which \eqref{eq:condition} holds is
\begin{equation}
R_{\min}=I(W)-\min_i \left(I\left(W_i\right)\sqrt{\frac{V_\lambda\left(W\right)}{2^\lambda V\left(W_i\right)}}\right)
\label{eq:rmin}
\end{equation}
Fig. \ref{fig:R_min} shows $R_{\min}$ for different values of $\lambda$ for the BIAWGNC. Fig. \ref{fig:R_min} suggests that $\lim_{SNR\rightarrow\infty}R_{\min}=1-1/2^\lambda$. This will be proved theoretically later. In this entire derivation, we ignore the requirement that $R_i\cdot N_1\in \mathbb{Z}$, since
this requirement becomes redundant as $N_1$ increases. For the same reason, we also ignore the correction term.
\begin{figure}
\begin{center}
\includegraphics[width=0.5\columnwidth]{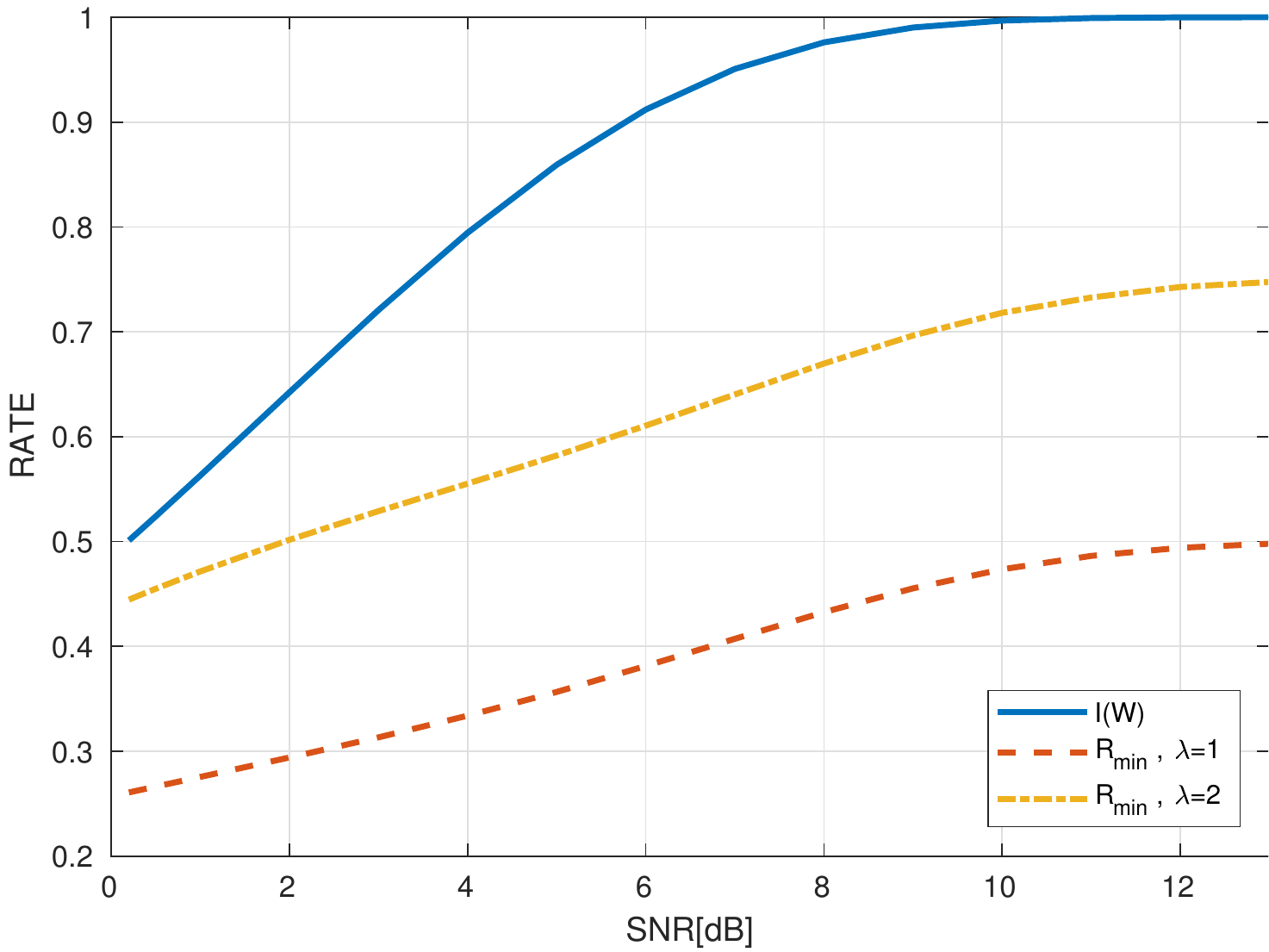}
\caption{The minimal rate for which \protect\eqref{eq:condition} holds compared to the channel capacity for the BIAWGNC}
\label{fig:R_min}
\end{center}
\end{figure}
This is compared to a naive approach where we simply use $2^{\lambda}$ 
codewords with blocklength $N_1=N/2^{\lambda}$ (without using the polar 
transformation at all). In this approach,
\begin{equation}
P_e\approx 2^\lambda Q\left([I(W)-R]\sqrt{\frac{N_1}{V(W)}}\right)=2^\lambda Q\left([I(W)-R]\sqrt{\frac{N}{V_{n,\lambda}(W)}}\right)
\label{eq:Pe_approx_naive}
\end{equation}
where $V_{n,\lambda}(W)\triangleq 2^\lambda V(W)$. Showing that $V_{\lambda}(W)\le V_{n,\lambda}(W)$ would mean that polarization is helpful. As can be seen in Fig. \ref{fig:dispersion}, for codes of rate $R>R_{\min}$, transmitted over the BIAWGNC, $V_{\lambda}(W)< V_{n,\lambda}(W)$, so polarization helps. For high SNR, Fig. \ref{fig:dispersion} suggests that $V_{\lambda}(W)\approx V(W)$.
\begin{figure}
	\begin{center}
		\includegraphics[width=0.5\columnwidth]{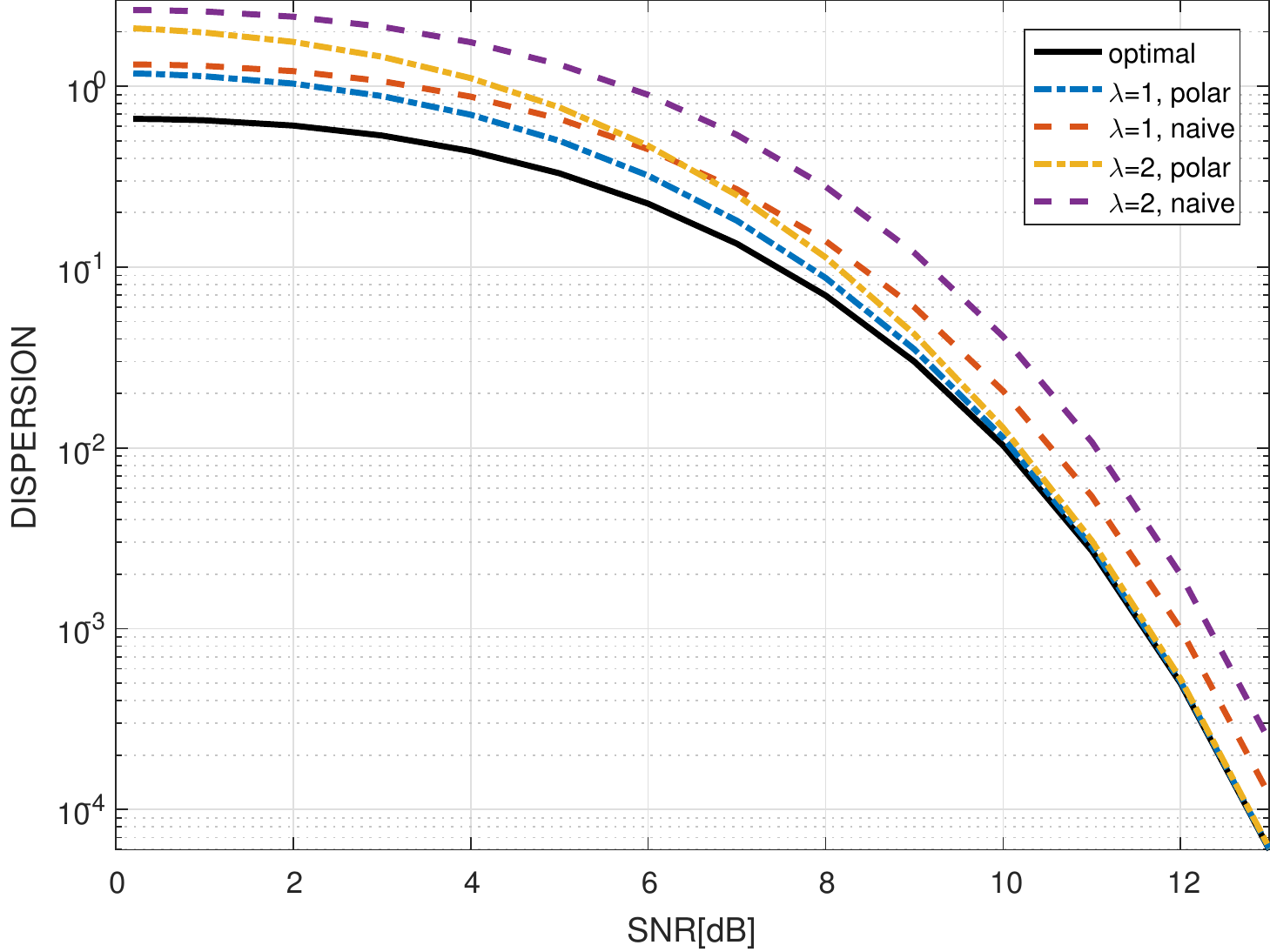}
		\caption{$V_\lambda(W)$ and $V_{n,\lambda}(W)$ for a BIAWGNC}
		\label{fig:dispersion}
	\end{center}
\end{figure}
This can be proved theoretically, but it requires the following lemma first. 
\begin{lemma} \label{lem:V_induction}
If $R$ is sufficiently large, s.t. \eqref{eq:condition} holds, i.e. $R>R_{\min}$, then, for $\lambda\ge 1$, $V_\lambda(W)$ can be calculated recursively using 
$
V_\lambda(W)=0.5\left(\sqrt{V_{\lambda-1}\left(W^-\right)}+\sqrt{V_{\lambda-1}\left(W^+\right)}\right)^2
$
where $V_0(W)\triangleq V(W)$.
\end{lemma}	
\begin{IEEEproof}
%
For $\lambda>1$, note that $W_i$ are sub-channels of $W^-$ for $i=0,\dots,2^{\lambda-1}-1$, and sub-channels of $W^+$ for $i=2^{\lambda-1},\dots,2^\lambda-1$.
\begin{align}
V_\lambda(W)
&=0.5\left(\frac{\sum_{i=0}^{2^{\lambda-1}-1}\sqrt{V\left(W_i\right)}}{\sqrt{2^{\lambda-1}}}+\frac{\sum_{i=2^{\lambda-1}}^{2^{\lambda}-1}\sqrt{V\left(W_i\right)}}{\sqrt{2^{\lambda-1}}}\right)^2\\
&=0.5\left(\sqrt{V_{\lambda-1}\left(W^-\right)}+\sqrt{V_{\lambda-1}\left(W^+\right)}\right)^2
\end{align}
where the first equality follows from rewriting \eqref{eq:Vp_def}, and the second one follows from applying \eqref{eq:Vp_def} for $\lambda-1$ instead of $\lambda$.
\end{IEEEproof}
We now consider the case of $I(W)$ close to $1$ (the high SNR case) and claim the following.
\begin{lemma}
Consider a BMS channel $W$ and suppose that
$V(W)$ can be linearly approximated as\footnote{$o(\delta)$ denotes a term which is negligible in its absolute value compared to $\delta$ for sufficiently small $\delta$} $V(W)=\alpha\delta(W)+o(\delta(W))$ where $\alpha$ is some constant and $\delta(W) \defined 1-I(W)$.
Also assume that $W^+$ and $W^-$ satisfy the same property (with the same value of $\alpha$).
Then $V_\lambda(W)=V(W)+o\left(\delta(W)\right)$ for $\lambda\ge 1$. Furthermore, under our assumptions, $\lim_{I(W)\rightarrow 1} R_{\min}=1-1/2^\lambda$.
\label{lem:dispersion}
\end{lemma}
We prove the Lemma in Appendix \ref{app:lem_dispersion_proof}

The conditions of Lemma \ref{lem:dispersion} hold in particular for the BEC.
The condition regarding $W$ holds with $\alpha=1$ since $V(W)=I(W)\left[1-I(W)\right]$. The conditions regarding $W^+$ and $W^-$ hold since by \cite[Proposition 6]{arikan2009channel}, if $W$ is a BEC, so are $W^+$ and $W^-$.
For the BIAWGNC, \cite[Fig. 4]{guillen2013extremes} suggests that the condition on $W$ is valid. Furthermore, if $W$ is a BIAWGNC, then $W^+$ and $W^-$ can be approximated as BIAWGNCs as well \cite{trifonov2011generalized}. These arguments suggest that the conditions of Lemma \ref{lem:dispersion} are also valid for the BIAWGNC. Fig. \ref{fig:dispersion} supports this conjecture. 

Consider transmission over a BMS channel $W$ with code rate $R>R_{\min}$, where $I(W)$ is sufficiently close to $1$. One could think that under the conditions of Lemma \ref{lem:dispersion} on $W$, $W^-$ and $W^+$, and the normal approximation to the error probability, the best achievable error probability of the polar concatenated coding scheme, \eqref{eq:Pe_approx_polar}, is larger only by a factor of $2^\lambda$ compared to the best achievable error probability of an arbitrary code with the same total blocklength and rate, i.e. $Q(x/\sqrt{V(W)}) / Q(x/\sqrt{V_\lambda(W)})\approx 1$  for $x\triangleq\sqrt{N}\left[I(W)-R\right]$ and $I(W)$ close to $1$.

Unfortunately, this is \emph{not} true. Since $Q(u)\sim \exp\left(-0.5u^2\right)$ for large $u$, and, according to the conditions and claims of Lemma \ref{lem:dispersion}, $\lim_{I(W)\rightarrow 1}V(W),V_\lambda(W)= 0$, we have
\begin{equation}
\left.Q\left(x/\sqrt{V(W)}\right)\right/Q\left(x/\sqrt{V_\lambda(W)}\right)\approx \exp\left[-\frac{x^2}{2}\left(\frac{1}{V(W)}-\frac{1}{V_\lambda(W)}\right)\right] \label{eq:condition_q}
\end{equation}
for $I(W)$ sufficiently close to $1$.
The condition $$\lim_{I(W)\rightarrow 1} \frac{1}{V(W)}-\frac{1}{V_\lambda(W)}=0$$ is required for claiming that the RHS of \eqref{eq:condition_q} approaches one for $I(W)\rightarrow 1$. This condition is stronger than the one proved in Lemma \ref{lem:dispersion}, and it does not necessarily hold, since $1/V(W)-1/V_\lambda(W)=\frac{o(\delta)}{\alpha^2\delta^2+\alpha o\left(\delta^2\right)}$ does not necessarily approach zero for $\delta\rightarrow 0$.
I removed the example illustrated in the previous Fig. 6 in order to focus on our main contributions.
As an example, it is easy to prove using Lemma \ref{lem:V_induction}, that for the BEC and $I(W)=1-\delta$, $V(W)=\delta(1-\delta)$, but $V_1(W)=\delta +\Theta\left(\delta^{1.5}\right)$, so $1/V(W)-1/V_1(W)=\Theta\left(\delta^{-0.5}\right)$ and the RHS of \eqref{eq:condition_q} approaches zero for $I(W)\rightarrow 1$.
\rem{
\begin{figure}
  \begin{center}
    \includegraphics[width=0.5\columnwidth]{dispersion_gap.pdf}
    \caption{$1/V(W)-1/V_\lambda(W)$ for a BIAWGNC}
    \label{fig:dispersion_gap}
  \end{center}
\end{figure}
}

Suppose that the condition $R>R_{\min}$ holds. Then the error rate of the polar concatenated (naive) scheme is given by \eqref{eq:Pe_approx_polar} (\eqref{eq:Pe_approx_naive}, respectively). Hence, comparing $V_\lambda(W)$ to $V_{n,\lambda}(W)$, we can assess the usefulness of polarization compared to the naive approach with the same value of $\lambda$.
The following result shows that for large $N$, $V_\lambda(W)$ and $V_{n,\lambda}(W)$ can be used to compare between schemes with different values of $\lambda$.
Define $\hat{V}_\lambda(N,W)$ as the solution to 	
\begin{equation}
2^\lambda Q\left([I(W)-R]\sqrt{\frac{N}{V_\lambda(W)}}\right)= Q\left([I(W)-R]\sqrt{\frac{N}{\hat{V}_\lambda(N,W)}}\right) \label{eq:Qeq}
\end{equation}
for given $I(W)-R$, $\lambda$ and $N$. Also define $\hat{V}_{n,\lambda}(N,W)$ as the solution of 
\begin{equation}
2^\lambda Q\left([I(W)-R]\sqrt{\frac{N}{V_{n,\lambda}(W)}}\right)= Q\left([I(W)-R]\sqrt{\frac{N}{\hat{V}_{n,\lambda}(N,W)}}\right) \label{eq:Qneq}
\end{equation}
\begin{lemma}
$\hat{V}_\lambda(N,W)-V_\lambda(W)=\Theta\left(\frac{1}{N}\right)$, so $\lim_{N\rightarrow\infty}\hat{V}_\lambda(N,W)=V_\lambda(W)$. The same claim also holds for $\hat{V}_{n,\lambda}(N,W)$ and $V_{n,\lambda}(W)$.
\label{lem:hatV_proof}
\end{lemma}
We prove the lemma in Appendix \ref{app:lem_hatV_proof}.

For asymptotically large blocklengths the error probabilities predicted by Gallager's error exponents are more accurate (the error exponents are conjectured to be the correct ones under our sub-optimal decoding scheme). However, in the finite blocklength regime, the normal approximation is better. In our setting, we target the case of $N_1$ values which are not very large (e.g., of order 100). In the following section we show a very good match between the normal approximation and simulation results, using close to ML decoded, powerful outer algebraic codes. 

\subsection{Approximated converse for BIAWGNC}\label{subsec:convBIAWGN}
In \cite{shannon1959probability}, a lower bound on the optimal FER of optimal spherical codes over additive white Gaussian noise channels (AWGNCs) is provided. Since the BIAWGNC is a constrained version of a AWGNC with the same SNR, this bound can be treated as a lower bound on the FER for a BIAWGNC as well, i.e. for the same $\rm{SNR}=1/\sigma^2$, rate $R$ and block size $N$,  $P_{e,BIAWGNC}\left(\sigma,R,N\right)\ge P_{e,AWGNC}\left(\sigma,R,N\right)$. We obtain the following approximated lower bound on the achievable frame error rate after $\lambda$ polarization steps.
\begin{align}
P_e
&\ge
1-\prod_{i=0}^{2^\lambda-1}\left[1-P_e\left(W_i,R_i,N_1\right)\right] \label{eq:awgn_conv}\\
&\approx
1-\prod_{i=0}^{2^\lambda-1}\left[1-P_{e,BIAWGN}\left(\sigma_i,R_i,N_1\right)\right]\\
&\ge
1-\prod_{i=0}^{2^\lambda-1}\left[1-P_{e,AWGN}\left(\sigma_i,R_i,N_1\right)\right]\\
&\ge \min_{R_i}\left[1-\prod_{i=0}^{2^\lambda-1}\left[1-P_{e,AWGN}\left(\sigma_i,R_i,N_1\right)\right]\right] 
\end{align}
The first inequality is under the assumption that each outer code $A_i$ is ML decoded given the channel observations and the previously decoded codewords of the outer codes $A_0,\ldots,A_{i-1}$. We bound the error rate from below by assuming a genie aided decoder: the genie informs us what was the actual transmitted codeword of the outer code, $A_i$, immediately after we decode it (so that it can be used for decoding the codewords of the following outer codes, $A_{i+1},A_{i+2},\ldots$).
The approximation in the second line follows from the approximation of the sub-channels $W_i$ as BIAWGNCs, when $W$ is a BIAWGNC \cite[Equations (7)-(8)]{trifonov2011generalized}. The second inequality follows from the explanation above. The term $P_{e,AWGN}\left(\sigma,R,N\right)$ is calculated using \cite{shannon1959probability}
\begin{equation}
P_{e,AWGN}\left(\sigma,R,N\right)\approx\frac{1}{\sqrt{N\pi}}\frac{1}{\sqrt{1+G^2}\sin\theta}\cdot \frac{\left[G\sin\theta\exp\left(-\frac{1}{2\sigma^2}+\frac{G\cos\theta}{2\sigma}\right)\right]^N}{G\sin^2\theta/\sigma-\cos\theta}
\end{equation}
where $G=0.5\left(\cos\theta/\sigma+\sqrt{\cos^2\theta/\sigma^2+4}\right)$ and $\theta$ is computed by solving $2^{NR}\approx \frac{\sqrt{2\pi N}\sin\theta\cos\theta}{\sin^N\theta}$. The two approximations above are extremely accurate for $N\ge 100$.

\subsection{Converse for the BEC} \label{subsec:convBEC}
Combining \eqref{eq:awgn_conv} with \cite[Theorem 38]{polyanskiy2010channel} and the fact that the sub-channels of a BEC are also BEC, yields that for all polar concatenated codes over the BEC with capacity $I(W)$, $\lambda$ polarization steps, and blocklength $N_1$
\begin{align}
P_e
&\ge
1-\prod_{i=0}^{2^\lambda-1}\left[1-P_e\left(W_i,R_i,N_1\right)\right]\\
&\ge
1-\prod_{i=0}^{2^\lambda-1}\left[1-P_{c,BEC}\left(I\left(W_i\right),R_i,N_1\right)\right]\\
&\ge
\min_{R_i}\left[1-\prod_{i=0}^{2^\lambda-1}\left[1-P_{c,BEC}\left(I\left(W_i\right),R_i,N_1\right)\right]\right] \label{eq:converse}
\end{align}
where \begin{equation}
P_{c,BEC}\left(I(W),R,N\right)
\defined
\sum_{l=\lfloor N(1-R)\rfloor+1}^N\binom{N}{l}\left[1-I(W)\right]^lI(W)^{N-l}\left(1-2^{N(1-R)-l}\right)
\end{equation}

\section{Comparison with simulation results} \label{sec:simulation}
We start by comparing the actual performance of BCH-polar codes, for short blocklengths and one polarization step ($\lambda=1$) transmitted over a BIAWGNC, with the normal approximation-based expression \eqref{eq:V_approx}, for the best polar-concatenated code with these parameters using the dynamic programming algorithm described in Section \ref{sec:improved_bnds_and_approx}. The total code rate was $R=1/2$. We used two setups. In the first, $N_1=64$ and $N=128$. In the second setup, $N_1=128$ and $N=256$. We also calculated the normal approximation \eqref{eq:Q_biawgn} to the best achievable error probability when using a $(128,64)$ ($(256,128)$) code. For each SNR, the outer code rates that minimize \eqref{eq:V_approx} were calculated as a by-product of the dynamic programming algorithm, and are shown in Tables \ref{tab:N=128} and \ref{tab:N=256}.
\begin{table}
	\caption{Rates of outer codes in the $\left(\lambda,N_1,R\right)=(1,64,1/2)$ scheme for the BIAWGNC}
	\label{tab:N=128}
	\begin{center}
		\begin{tabular}{ccc}
			SNR[dB] &$N_1\cdot R_0$& $N_1\cdot R_1$\\
			1 &17 &47\\
			1.5 &16&48\\
			2 &16&48\\
			2.5 &16&48\\
			3 &15&49\\
			3.5 &15&49\\
			4 &14&50
		\end{tabular}
	\end{center}\label{table:N=128}
\end{table}
\begin{table}
	\caption{Rates of outer codes in the $\left(\lambda,N_1,R\right)=(1,128,1/2)$ scheme for the BIAWGNC}
	\label{tab:N=256}
	\begin{center}
		\begin{tabular}{ccc}
			SNR[dB] &$N_1\cdot R_0$& $N_1\cdot R_1$\\
			0.5 &34 &94\\
			1 &33&95\\
			1.5 &33&95\\
			2 &32&96\\
			2.5 &31&97\\
			3 &31&97\\
			3.5 &30&98
		\end{tabular}
	\end{center}
\end{table}

Due to the results in Table \ref{tab:N=128}, we used (64,18,22) and (64,45,8) extended BCH codes, whose generator matrices appear in \cite{bch_matrices}, as outer codes in the simulated BCH-polar coding scheme for $N=128$, and decoded them using OSDs of order 5 \cite{bch_matrices}. The total rate of the scheme is $R=63/128$, which is close to the planned rate. The (128,64,22) BCH code was decoded using OSD of order 5 as well. The consumed processor time (measured in clock ticks) normalized by the number of information bits of the decoders of BCH-polar and BCH codes were compared as well. The error rates and decoding times of the various schemes are shown in Figs. \ref{fig:polarBCH} and \ref{fig:polarBCH128time}. Fig. \ref{fig:polarBCH} suggests that the normal approximation we use is accurate in the SNR range examined. We also see that the BCH-polar code suffers a loss of 0.7dB compared to the BCH code (Fig. \ref{fig:polarBCH}), but is about 25 to 1000 times faster, depending on the SNR (Fig. \ref{fig:polarBCH128time}). Compared to the (128,64) list SC decoder with CRC (with list size 32), whose FER results were taken from \cite{liva2016survey}, the BCH-polar code suffers a loss of only 0.25dB.

For extended BCH with $N=128$ and $R=1/2$ we see in Fig. \ref{fig:polarBCH} that the normal approximation is accurate for low SNR, but for high SNR it slightly underestimates the FER. We conjecture the same behavior for other blocklengths and rates. Therefore, using \eqref{eq:V_approx}
to estimate the optimal rates for BCH-polar codes would be accurate for low SNR (considering that not all BCH rates are feasible), but for high SNR it would overestimate the rates of the good sub-channels. That's why we pick slightly lower $R_1$ and slightly higher $R_0$ than the ones in Table \ref{tab:N=128} while designing the BCH-polar codes.

The results for $N=256$ show a similar trend. Due to the rates in Table \ref{tab:N=256}, and since $R_1$ should be slightly lowered (and $R_0$ slightly increased) the chosen extended BCH codes were (128,36,32) and (128,92,12), whose generator matrices appear in \cite{bch_matrices}.
For comparison, we have taken the FER results of the (255,131) BCH code from \cite{performance2016wonterghem}, and measured the average processing time of the OSD of order 5 of this code. 
This time, the BCH-polar code suffers a loss of only $0.5{\rm dB}$ compared to the BCH code (Fig. \ref{fig:polarBCH}), but is about 1000 times faster (Fig. \ref{fig:polarBCH128time}).
Figs. \ref{fig:polarBCH} and \ref{fig:polarBCH128time} also show the following:
\begin{enumerate}
\item For ${\rm SNR}<2{\rm dB}$, the (256,128) BCH-polar code has a higher FER compared to the (128,64) BCH code, but it requires a lower processing time per information bit. 
\item For $2{\rm dB}<{\rm SNR}<3{\rm dB}$, the (256,128) BCH-polar code has a lower FER and lower processing time per information bit compared to the (128,64) BCH code.
\item For ${\rm SNR}>3{\rm dB}$, the (256,128) BCH-polar code has a lower FER than the (128,64) BCH code, but it requires higher processing time per information bits.
\end{enumerate}
Note that when decoding a (256,128) BCH-polar code, we need to decode two BCH codes of blocklength 128. However, the rate of the first is lower than $1/2$ while the rate of the second is higher than $1/2$. Also, the required complexity of the OSD algorithm is maximal for rate $1/2$ codes. Therefore, it is usually more efficient to decode a (256,128) BCH-polar code compared to the decoding of two (128,64) BCH codes. However, the decoding of the (256,128) BCH-polar code also requires some handling of soft information due to the polar transformation. We did not attempt to implement this part in our algorithm efficiently.

\begin{figure}
	\begin{center}
		\includegraphics[width=0.5\columnwidth]{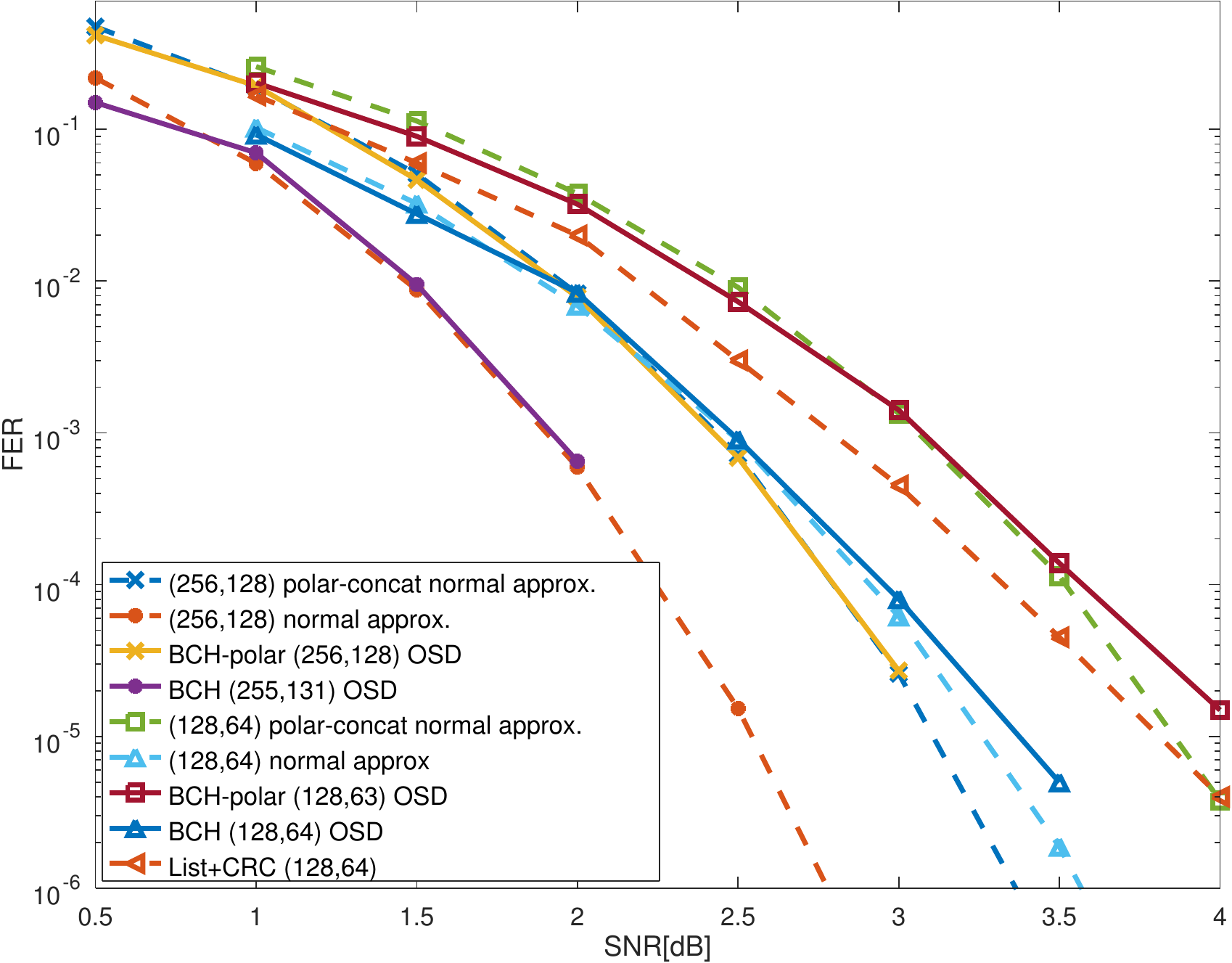}
		\caption{FER normal approximations and simulation results for BCH-polar codes with $\lambda=1$, and BCH codes with different blocklengths and rates  close to 1/2 for the BIAWGNC, compared with (128,64) List decoder with CRC from \cite{liva2016survey}}
		\label{fig:polarBCH}
	\end{center}
\end{figure}
\begin{figure}
	\begin{center}
		\includegraphics[width=0.5\columnwidth]{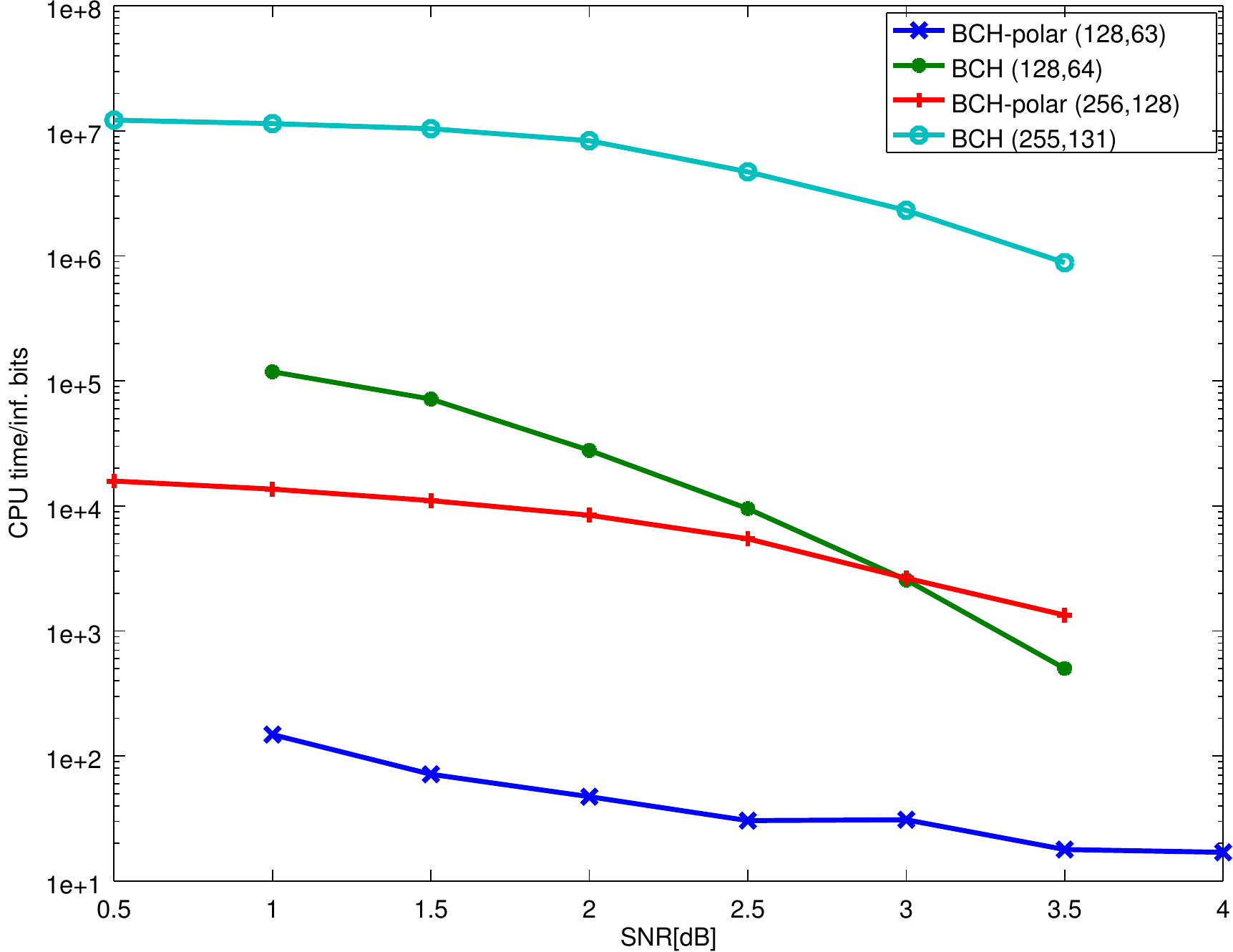}
		\caption{Comparison of processor time per information bits of BCH-polar codes with $\lambda=1$, and BCH codes with different lengths and rate close to 1/2 for the BIAWGNC.}
		\label{fig:polarBCH128time}
	\end{center}
\end{figure}
We also compared our results to simulation results from the literature. First, we considered the setup in \cite[Section VIII.A]{wang2016interleaved}, where $\lambda=3$, $N_1=128$, $N=1024$, and the channel is a BEC. The overall rate of the code is $R=0.4$.
We calculated the achievable upper bound on the FER, with typical random (the expurgated random ensemble) outer codes, using the recursive algorithm from Lemma \ref{lem:exact_recursion}.
We then computed an upper bound on the achievable FER by finding the best rate division between sub-codes given the total code rate $2^\lambda R$, that brings the RHS of \eqref{eq:theo37} to a minimum as described in Section \ref{sec:improved_achievable_BEC}.
Our upper bounds were computed twice. Once by optimizing the rate division for a fixed BEC, $W$, with erasure rate $0.4$ as in \cite{wang2016interleaved}, and once by optimizing the rate division for the actual BEC we are transmitting over, for each point in the graph. The corresponding graphs are denoted by ``BEC(0.4)'' and ``opt.'' in Fig. \ref{fig:BEC_FER}. We have also plotted the normal approximation \eqref{eq:V_approx} to the best achievable error probability, and the converse bound in Section \ref{subsec:convBEC}.
\begin{figure}
\begin{center}
\includegraphics[width=0.5\columnwidth]{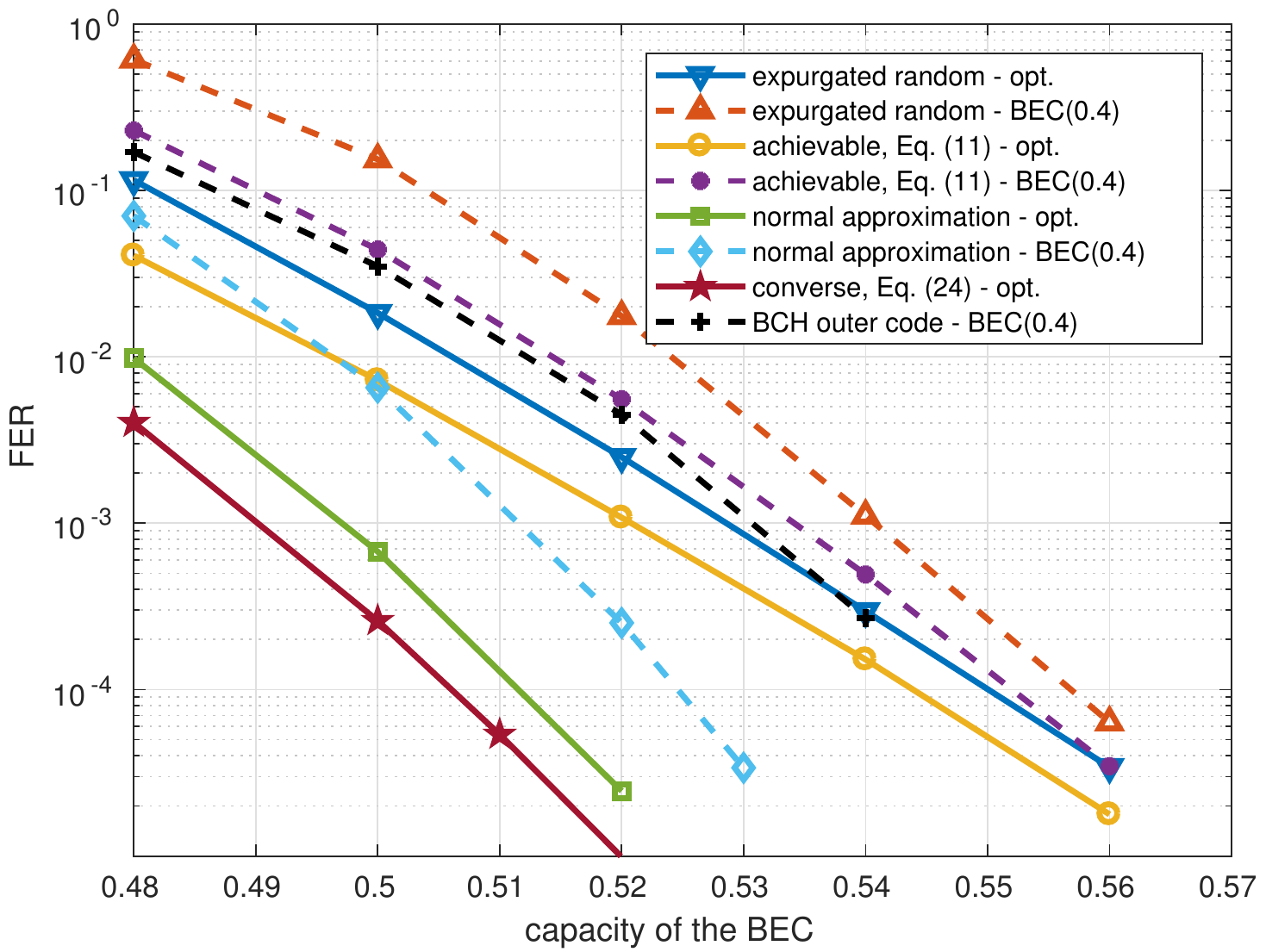}
\caption{FER upper bounds for concatenated polar codes with outer code length $N_1=128$ and inner polar code length 8, compared with BCH-polar code results from \cite[Fig. 5]{wang2016interleaved}.}
\label{fig:BEC_FER}
\end{center}
\end{figure}

The graphs show small gaps between the achievable bound, Equation \eqref{eq:theo37}, and the actual results with BCH codes. The normal approximation \eqref{eq:V_approx} has a somewhat lower FER, while the bound based on Lemma \ref{lem:exact_recursion} is less tight. The converse \eqref{eq:converse} is close to the normal approximation.
For comparison, note that a standard SC decoded polar code of length $N=1024$, yields $\rm{FER}\approx 2\cdot 10^{-3}$ for $I(W)=0.6$ \cite{wang2016interleaved}.

\comment{\color{blue}The outer-code rates obtained while optimizing the plotted bounds and approximations are shown in Table \ref{tab:rates}.
\begin{table}
\caption{Rates of random outer codes in the $\left(\lambda,N_1,R\right)=(3,128,0.4)$ scheme for the BEC channel}
\label{tab:rates}
\begin{center}
\begin{tabular}{ccc}
Capacity & Ensemble & $N_1\cdot$Rates\\
0.48& typical &0 4 10 68 22 86 98 121\\
0.48& achievable &0 3 10 68 23 87 98 120\\
0.48& normal approx. &0 3 9 68 21 86 98 124\\
0.5& typical &0 3 10 69 23 87 98 119\\
0.5& achievable &0 2 10 70 23 88 98 118\\
0.5& normal approx. &0 2 8 68 21 87 99 124\\
0.52& typical,achievable &0 2 10 71 24 88 98 116\\
0.52 & normal approx. &0 0 7 69 21 88 99 125\\
0.54& typical &0 1 10 72 25 89 98 114\\
0.54& achievable &0 1 10 73 25 89 98 113\\
0.56& typical &0 0 10 74 26 89 98 112\\
0.56& achievable &0 1 10 74 27 89 97 111\\
0.6& typical,achievable &0 0 12 75 29 89 95 109\\
0.6& normal approx. &0 0 1 73 18 91 101 125\\
0.6& BCH-polar \cite{wang2016interleaved} &0 0 0 71 8 99 113 120
\end{tabular}
\end{center}
\end{table}
We observed that the rates of the outer codes in our scheme are different from the rates in \cite{wang2016interleaved}, but the rates obtained by the normal approximation are the closest to the BCH-polar code}
	
The second setup we considered is taken from \cite[Section IV]{trifonov2011generalized}, where $\lambda=3$, $N_1=127$, $N=1016$, and the channel is a BIAWGNC. The overall rate of the code is $R=1/2$.
We first calculated the achievable upper bound on the FER from Lemma \ref{lem:exact_recursion}. Then we calculated the normal approximation \eqref{eq:V_approx} to the best achievable error probability. Both graphs were obtained by optimizing over the best rate split for each SNR point in the graph. We also plot the normal approximation given optimal rate division for fixed $\rm{SNR}=3dB$.
We compared these bounds with a simulation of the BCH-polar code in \cite{trifonov2011generalized} using an outer OSD of order 5. Note that in \cite{trifonov2011generalized} the outer code rates were optimized only for $\rm{SNR}=3\rm{dB}$. As can be seen, the normal approximation is close to the performance of the scheme with outer BCH codes and almost not affected by using a fixed rate division.
The figure also shows the performance of a standard polar code with $N=1024$ and $R=1/2$. Finally, we have plotted the converse for polar concatenated scheme as was described in Section \ref{subsec:convBIAWGN}, Equation \eqref{eq:awgn_conv}. We see that for a desired FER of $\approx 2\cdot 10^{-4}$ the BCH-polar code is only 0.75dB worse than the converse, so under the given $\lambda$, $N_1$ and $R$, we cannot gain more than 0.75dB by smartly choosing the outer codes. Finally note that the normal approximation for the best (1016,508) code at SNR = 1.25dB shows only 1dB improvement compared to the BCH-polar code.
\begin{figure}
\begin{center}
\includegraphics[width=0.5\columnwidth]{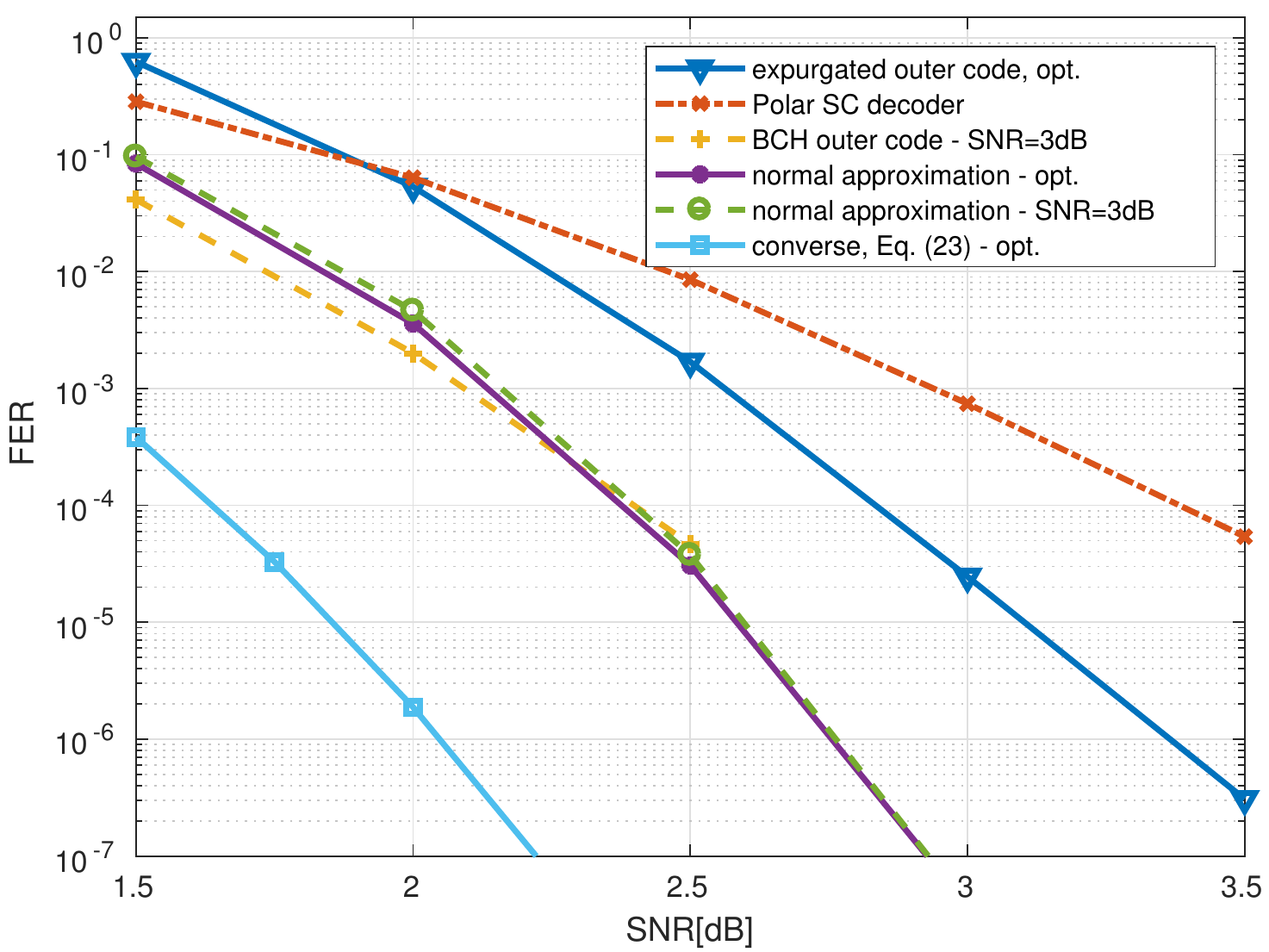}
\caption{FER upper bounds for concatenated polar codes with outer random code length $N_1=127$, inner polar code length 8, compared with BCH-polar code from \cite{trifonov2011generalized}}
\label{fig:AWGN_FER}
\end{center}
\end{figure}
\section{Conclusion}\label{sec:conclusion}
We studied the properties of a concatenated scheme of polar codes with good outer codes. We obtained an upper bound on the error exponent using the corresponding expurgated random coding ensemble, and calculated a lower bound on the error exponent, which is valid for all channels with a given capacity. We obtained converse and approximated converse results, as well as a dispersion-based normal approximation to the performance for finite length codes, which can also be used to determine the required rate split between the outer codes. We showed good agreement between this prediction and simulation results for BCH-polar codes, when transmitting over the BIAWGNC.

\appendices
\section{Proof of Lemma \ref{lem:f_property}} \label{app:lem_f_property_proof}
We first prove by induction that $E_\lambda(W,R)$ is finite for all integer $\lambda$ and $R>0$.
$E_{\lambda}(W,R)$ is finite for $\lambda=0$ and $R>0$, since both $E_r(W,R)$ and $E_{ex}(W,R)$ are finite: $E_r(W,R)\le E_0(W,1)$, and for binary-input channels (except for a perfect channel for which the two inputs can never be confused at the receiver), $E_{ex}(W,R)$ is finite for $R>0$, which can be derived from \cite[Equations (5.7.14)-(5.7.16)]{galbook}.
Now, assume $E_{\lambda-1}(W,R)$ is a finite function of $R$ for $R>0$. Then, $E_{m,\lambda}(W,R,R_1)$ defined in \eqref{eq:Eml_WRR1} is finite, since $E_{\lambda-1}\left(W^-,R_1\right)$ can be infinite only for $R_1=0$, but then $2R-R_1=2R>0$ and $E_{\lambda-1}\left(W^+,2R-R_1\right)$ is finite, so $E_{m,\lambda}(W,R,R_1)$ is finite too. This also means that $E_{\lambda}(W,R)=0.5\max_{R_1} E_{m,\lambda}(W,R,R_1)$ is finite. This shows that $E_{\lambda}\left(W,R\right)$ is finite for all integer $\lambda$ and $R>0$.
Note however that $E_\lambda(W,0)=\infty$ for all integer $\lambda$ and any channel $W$. This can be shown by induction, since for $R=0$ \eqref{eq:maxmin_recursion} yields $E_\lambda(W,0)=0.5\min\left[E_{\lambda-1}\left(W^+,0\right),E_{\lambda-1}\left(W^-,0\right)\right]$ and $E_\lambda(W,0)=\infty$ for $\lambda=0$ due to \eqref{eq:E_r_def}.
	
Next, we prove by induction that $E_{\lambda}(W,R)$ is decreasing in $R$. The claim is trivial for $\lambda=0$ since $E_{0}(W,R)=E(W,R)$ is defined in \eqref{eq:E(W,R)def} as a maximum between two continuous, decreasing functions. Hence it is decreasing whether it equals $E_{ex}(W,R)$ or $E_r(W,R)$.
Now assume that $E_{\lambda-1}(W,R)$ is decreasing in $R$ for any $W$.
First consider rates $R$ for which
\begin{equation}
\left\{\begin{array}{l}
E_{\lambda-1}\left(W^-,0^+\right)\ge E_{\lambda-1}\left(W^+,2R\right)\\
E_{\lambda-1}\left(W^-,2R\right)\le E_{\lambda-1}\left(W^+,0^+\right)
\end{array}\right. \label{eq:mid_rates}
\end{equation}
(for rates $R>1$ we define for all $\lambda$, $E_\lambda(W,R)\equiv 0$). This case is depicted in Fig. \ref{fig:required}. By \eqref{eq:maxmin_recursion} and Fig. \ref{fig:required}, $2E_{\lambda}(W,R)$ is the height of the intersection point of $E_{\lambda-1}\left(W^-,R_1\right)$, which is a decreasing function of $R_1$, and $E_{\lambda-1}\left(W^+,2R-R_1\right)$, which is an increasing function of $R_1$. Therefore, increasing $R$ would move $E_{\lambda-1}\left(W^+,2R-R_1\right)$ to the right, as can be seen in Fig. \ref{fig:required}, thus decreasing the intersection point height, $2E_{\lambda}(W,R)$, and increasing $\hat{R_1}=\argmax_{R_1} \min\left[E_{\lambda-1}\left(W^-,R_1\right),E_{\lambda-1}\left(W^+,2R-R_1\right)\right]$. This means $E_{\lambda}(W,R)$ is decreasing, and $\hat{R_1}=\hat{R_1}(R)$ is an increasing function of $R$, as can be seen in Fig. \ref{fig:required}: $\hat{R}_2\triangleq \hat{R}_1(R+\Delta R)>\hat{R}_1(R)$ for $\Delta R>0$.

If \eqref{eq:mid_rates} does not hold, i.e., $E_{\lambda-1}\left(W^-,0^+\right) < E_{\lambda-1}\left(W^+,2R\right)$ ($E_{\lambda-1}\left(W^+,0^+\right) < E_{\lambda-1}\left(W^-,2R\right)$, respectively), $\hat{R_1}=\max(0,2R-1)$ ($\hat{R_1}=\min(1,2R)$) and $E_\lambda\left(W,R\right)=0.5E_{\lambda-1}\left(W^-,\max(0^+,2R-1)\right)$ ($0.5E_{\lambda-1}\left(W^+,2R-\min(1,2R)\right)$). Trivially, $E_\lambda\left(W,R\right)$ is decreasing in this case as well.
Thus we have shown that $E_{\lambda}(W,R)$ is decreasing in $R$ for all integer $\lambda$, and that $\hat{R_1}$ is an increasing (but not strictly increasing) function of $R$.

We proceed by proving by induction that $E_{\lambda}(W,R)$ is convex.
The claim is trivial for $\lambda=0$ since $E_{0}(W,R)=E(W,R)$ is defined in \eqref{eq:E(W,R)def} as a maximum between two continuous, convex functions, $E_r(W,R)$ and $E_{ex}(W,R)$, and since maximization preserves convexity.
Now assume that $E_{\lambda-1}(W,R)$ is convex in $R$ for all BMS channels $W$. Since it was already shown that $E_{\lambda-1}(W,R)$ is also decreasing for all BMS channels $W$, it follows that $\left|\partial E_{\lambda-1}\left(W^-,r\right)/\partial r\right|_{r=\hat{R_1}(R)}$ is decreasing in $R$ (recalling that $\hat{R_1}(R)$ is an increasing function of $R$).
Since $E_{\lambda-1}\left(W^+,2R-\hat{R_1}\right) = 2E_{\lambda}(W,R)$, and $E_{\lambda}(W,R)$ is a decreasing function of $R$, $E_{\lambda-1}\left(W^+,2R-\hat{R}_1(R)\right)$ is also decreasing in $R$. But $E_{\lambda-1}\left(W^+,r\right)$ is decreasing in $r$. Thus $2R-\hat{R_1}(R)$ is increasing in $R$.
Since in addition $E_{\lambda-1}(W^+,r)$ is convex and decreasing, it follows that $\left|\partial E_{\lambda-1}\left(W^+,2R-r\right)/\partial r\right|_{r=\hat{R_1}(R)}$ is decreasing in $R$. 
We proceed by claiming that
\begin{equation}
\left|\partial E_{\lambda}\left(W,R\right)/\partial R\right|
=
\left(
\frac{1}{\left|\partial E_{\lambda-1}\left(W^-,r\right)/\partial r\right|_{r=\hat{R_1}(R)}} + \frac{1}{\left|\partial E_{\lambda-1}\left(W^+,2R-r\right)/\partial r\right|_{r=\hat{R_1}(R)}}
\right)^{-1}
\label{eq:triangle_deriv}
\end{equation}
Assume that \eqref{eq:triangle_deriv} holds. Since both $\left|\partial E_{\lambda-1}\left(W^-,r\right)/\partial r\right|_{r=\hat{R_1}(R)}$ and $\left|\partial E_{\lambda-1}\left(W^+,2R-r\right)/\partial r\right|_{r=\hat{R_1}(R)}$ decrease with $R$, it follows that $\left|\partial E_{\lambda}\left(W,R\right)/\partial R\right|$ is also decreasing in $R$. Since, in addition, $\partial E_{\lambda}\left(W,R\right)/\partial R < 0$ (due to the fact that $E_\lambda(W,R)$ is decreasing in $R$), this means $E_{\lambda}\left(W,R\right)$ is convex.

It remains to prove \eqref{eq:triangle_deriv}. First assume that \eqref{eq:mid_rates} holds. Consider Fig. \ref{fig:required}, which shows $E_{\lambda-1}\left(W^-,r\right)$, $E_{\lambda-1}\left(W^+,2R-r\right)$ and $E_{\lambda-1}\left(W^+,2(R+\Delta R)-r\right)$ ($\Delta R$ is small fixed value) as a function of $r$, together with intersection points standing for $2E_{\lambda}(W,R)$ and $2E_{\lambda}(W,R+\Delta R)$. Now, $2\Delta E \triangleq 2 E_\lambda(W,R) - 2E_\lambda(W,R+\Delta R)$ is the height of the triangle formed by the vertices $(\hat{R}_1,2E_{\lambda}(W,R))$, $(\hat{R}_2,2E_{\lambda}(W,R+\Delta R))$ and $(\hat{R}_1+2\Delta R,2E_{\lambda}(W,R))$ in Fig. \ref{fig:required}. For $\Delta R\rightarrow 0$, the slopes of this triangle's edges are approximated as
\begin{align}
a &\triangleq 2\Delta E/(\hat{R}_2-\hat{R}_1)\approx\left|\partial E_{\lambda-1}\left(W^-,r\right)/\partial r\right|_{r=\hat{R_1}}\\
b &\triangleq 2\Delta E/(\hat{R}_1+2\Delta R-\hat{R}_2)\approx\left|\partial E_{\lambda-1}\left(W^+,2R-r\right)/\partial r\right|_{r=\hat{R_1}}
\end{align}

It can be easily verified that $2\Delta E= 2\Delta R (1/a+1/b)^{-1}$, and together with $\Delta E/\Delta R\rightarrow \left|\partial E_{\lambda}\left(W,R\right)/\partial R\right|$, the approximations above yield \eqref{eq:triangle_deriv}. Now, if \eqref{eq:mid_rates} does not hold, then as was noted above, either
$E_\lambda\left(W,R\right) = 0.5E_{\lambda-1}\left(W^-,\max(0^+,2R-1)\right)$ or $E_\lambda\left(W,R\right) = 0.5E_{\lambda-1}\left(W^+,2R-\min(1,2R)\right)$.
In both cases $E_\lambda\left(W,R\right)$ is convex.
\begin{figure}
\begin{center}
\includegraphics[width=0.5\columnwidth]{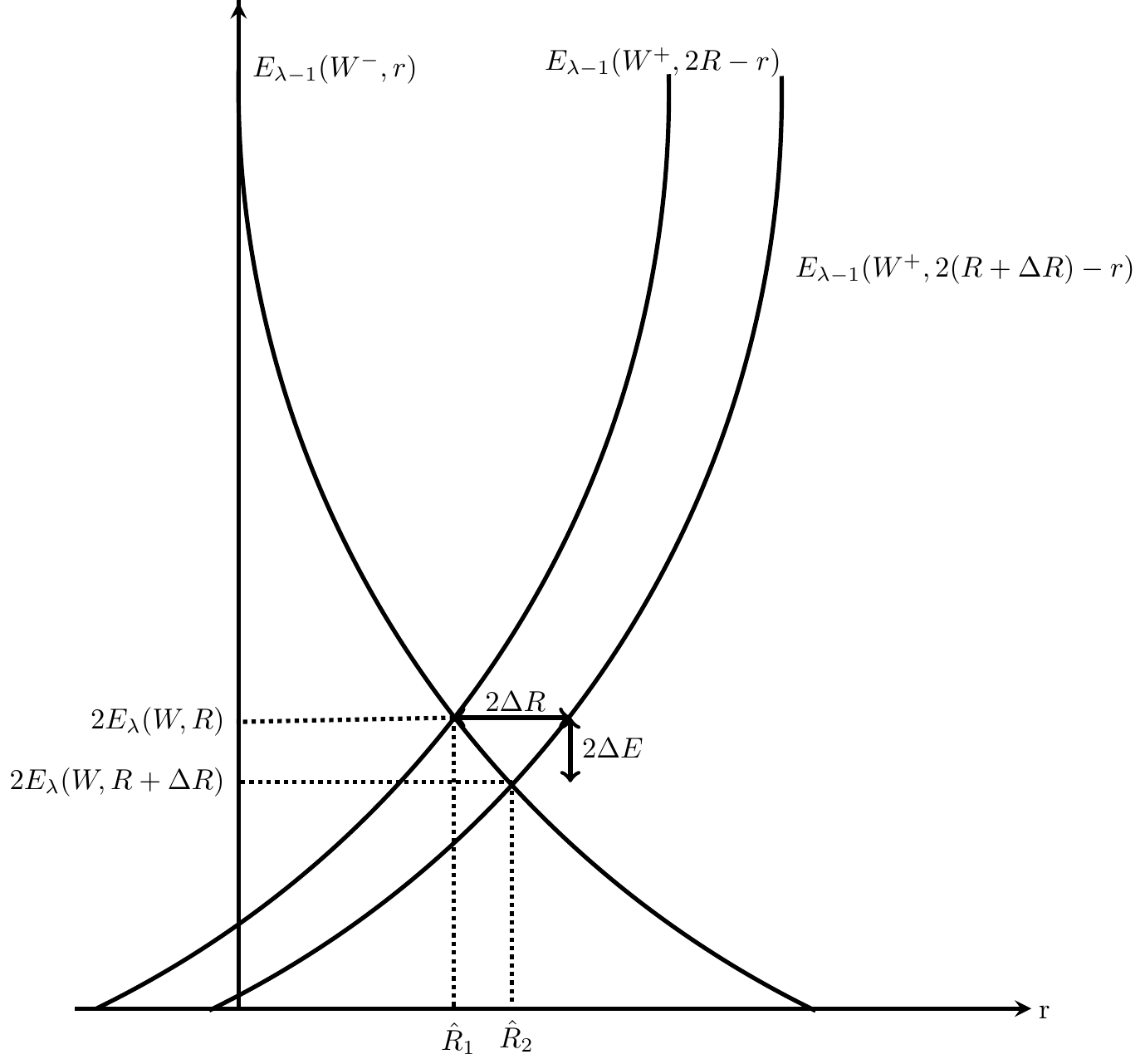}
\caption{Functions required to calculate $E_\lambda\left(W,R\right)$ and its derivative.}
\label{fig:required}
\end{center}
\end{figure}
Finally, the claim that for $R>0$, $\left|\partial E_{\lambda}\left(W,R\right)/\partial R\right|<\infty$ is due to the fact that $E_{\lambda}(W,R)$ is a finite, convex, decreasing function for all $R>0$.
\hfill\(\IEEEQEDhere\)

\section{Proof of Lemma \ref{lem:dispersion}} \label{app:lem_dispersion_proof}

We first prove by induction on $\lambda$ that $V_\lambda(W)=V(W)+o\left(\delta(W)\right)$.
This claim is trivial for $\lambda=0$. Now assume it holds for $\lambda-1$.
By the induction assumption,
\begin{equation}
V_{\lambda-1}(W) = V(W) + o\left(\delta(W)\right) 
= \alpha\delta(W) + o\left(\delta(W)\right) 
\end{equation}
Now, $\epsilon_l()$ and $\epsilon_h()$, defined in Section \ref{sec:background}, satisfy
\begin{equation}
\epsilon_l(I(W)) = \delta(W) + o(\delta(W)) \quad , \quad
\epsilon_h(I(W)) =\delta(W) - \delta(W)^2
\end{equation}
(the claim on $\epsilon_l(I(W))$ follows by a first order Taylor expansion of $\epsilon_l(1-\delta)$ for small $\delta$).
Hence, recalling Section \ref{sec:background},
\begin{equation}
I(W)-I\left(W^-\right) = I(W^+)-I\left(W\right) = \delta(W) + o\left(\delta(W)\right)
\end{equation}
so that
\begin{equation}
I(W^-) = 1-2\delta(W) + o\left(\delta(W)\right) \quad , \quad I(W^+) = 1 + o\left(\delta(W)\right)
\end{equation}
By the assumption of the theorem regarding $W^-$ and $W^+$,
\begin{equation}
V(W^-) = \alpha \delta(W^-) + o(\delta(W^-)) = 2\alpha \delta(W) + o(\delta(W)) \quad , \quad 
V(W^+) = \alpha \delta(W^+) + o(\delta(W^+)) = o(\delta(W))
\end{equation}
Applying the induction assumption to $W^-$ and $W^+$, we know that
\begin{equation}
V_{\lambda-1}\left(W^-\right) = V(W^-) + o(\delta(W^-)) = 2\alpha\delta(W) + o\left(\delta(W)\right)
\: , \:
V_{\lambda-1}\left(W^+\right) = V(W^+) + o(\delta(W^+)) = o\left(\delta(W)\right)
\end{equation}
By Lemma \ref{lem:V_induction},
\begin{equation}
V_\lambda(W)=0.5\left(\sqrt{2\alpha\delta(W)+o\left(\delta(W)\right)}+\sqrt{o\left(\delta(W)\right)}\right)^2=\alpha\delta(W)+o\left(\delta(W)\right)=V(W)+o\left(\delta(W)\right)
\end{equation}
This concludes the proof of the first part of the lemma. 

We proceed by proving by induction on $\lambda$ that after $2^\lambda$ polarization steps of the channel $W$,
\begin{equation}
\label{eq:IW0}
I\left(W_0\right) = I\left(W^{-\dots -}\right) = 1 - 2^\lambda\delta(W) + o(\delta(W))
\end{equation}
and
\begin{equation}
\label{eq:IWi}
I\left(W_i\right) = 1 + o(\delta(W)) \qquad i\ne 0
\end{equation}
This claim is trivial for $\lambda=0$. Assuming it holds for $\lambda-1$, we note that if $I\left(W_i\right)=1+o(\delta(W))$, then, using essentially the same arguments that were used in the beginning of the proof
\begin{equation}
I\left(W_i\right)-I\left(W_i^-\right)=I\left(W_i^+\right)-I\left(W_i\right)=o(\delta(W)) \quad i\ne 0
\end{equation}
Therefore, for $i\ne 0$,
\begin{equation}
I\left(W_i^-\right) = 1+o(\delta(W)) \quad , \quad 
I\left(W_i^+\right) = 1+o(\delta(W))
\end{equation}
In addition, by the induction assumption,
\begin{equation}
I\left(W_0\right)-I\left(W_0^-\right) = I\left(W_0^+\right)-I\left(W_0\right) = 
2^{\lambda-1}\delta(W) + o(\delta(W))
\end{equation}
Hence, reapplying the induction assumption,
\begin{equation}
I\left(W_0^-\right) = 1-2^{\lambda}\delta(W) + o(\delta(W)) \quad , \quad
I\left(W_0^+\right) = 1+o(\delta(W))
\end{equation}
This concludes the proof of \eqref{eq:IW0} and \eqref{eq:IWi}.
Thus, after $\lambda$ polarization steps, $V\left(W_0\right)=2^\lambda\alpha\delta(W)+o(\delta(W))$ and $V\left(W_i\right)=o(\delta(W))$ for $i\ne 0$.
We conclude that
\begin{equation}
\lim_{I(W)\rightarrow 1}I\left(W_0\right)\sqrt{\frac{V_\lambda\left(W\right)}{2^\lambda V\left(W_0\right)}} =
\lim_{\delta(W)\rightarrow 0}
\left[1-2^\lambda\delta(W)+o(\delta(W))\right]
\sqrt{\frac{\alpha\delta(W)+o(\delta(W))}{2^{2\lambda} \alpha\delta(W)+o(\delta(W))}} =
1/2^\lambda
\end{equation}
and for $i\ne 0$,
\begin{equation}
\lim_{I(W)\rightarrow 1}I\left(W_i\right)\sqrt{\frac{V_\lambda\left(W\right)}{2^\lambda V\left(W_i\right)}} =
\lim_{\delta(W)\rightarrow 0}
\left[1+o(\delta(W))\right]
\sqrt{\frac{\alpha\delta(W)+o(\delta(W))}{2^{\lambda} o(\delta(W))}} = \infty
\end{equation}
Therefore,
\begin{equation}
\lim_{I(W)\rightarrow 1}\min_i \left(I\left(W_i\right)\sqrt{\frac{V_\lambda\left(W\right)}{2^\lambda V\left(W_i\right)}}\right)=1/2^\lambda
\end{equation}
Thus, by \eqref{eq:rmin}, $\lim_{I(W)\rightarrow 1}R_{\min}=1-1/2^\lambda$.
\hfill\(\IEEEQEDhere\)

\section{Proof of Lemma \ref{lem:hatV_proof}} \label{app:lem_hatV_proof}
For shortness of notation, define $x\triangleq I(W)-R$. Also denote $\hat{V}=\hat{V}_\lambda(N,W)$ and $V=V_\lambda(W)$. First, we will find a sufficient condition for
\begin{equation}
2^\lambda Q\left(x\sqrt{\frac{N}{V}}\right)\le Q\left(x\sqrt{\frac{N}{\hat{V}}}\right) \label{eq:Q}
\end{equation}
Recall that for $u>0$ the complementary Gaussian cumulative distribution function, $Q()$, satisfies,
\begin{equation}
\frac{1}{\sqrt{2\pi}u}\left(1-\frac{1}{u^2}\right)e^{-u^2/2}
\le Q(u) \le
\frac{1}{\sqrt{2\pi}u}e^{-u^2/2}
\label{eq:Qineqs}
\end{equation}
The left hand side (LHS) of \eqref{eq:Qineqs} implies that
$$
\frac{\sqrt{\hat{V}}}{x\sqrt{2\pi N}}\left(1-\frac{\hat{V}}{x^2N}\right)e^{-\frac{x^2N}{2\hat{V}}}\le Q\left(x\sqrt{\frac{N}{\hat{V}}}\right)
$$
while the RHS of \eqref{eq:Qineqs} implies that
$$
2^\lambda Q\left(x\sqrt{\frac{N}{V}}\right)\le \frac{2^\lambda\sqrt{V}}{x\sqrt{2\pi N}}e^{-\frac{x^2N}{2V}}
$$
Therefore, 
\begin{equation}
\sqrt{\hat{V}}\left(1-\frac{\hat{V}}{x^2N}\right)e^{-\frac{x^2N}{2\hat{V}}}\ge 2^\lambda \sqrt{V}e^{-\frac{x^2N}{2V}}
\label{eq:Qcond}
\end{equation}
is a sufficient condition for \eqref{eq:Q}. But \eqref{eq:Qcond} is equivalent to 
\begin{equation}
\frac{\ln \hat{V}}{2} + \ln\left(1-\frac{\hat{V}}{x^2N}\right)-\frac{x^2N}{2\hat{V}}\ge \lambda\ln 2 + \frac{\ln V}{2} - \frac{x^2N}{2V}
\label{eq:Qcond1}
\end{equation}
Since
$$
\ln\left(1-\frac{\hat{V}}{x^2N}\right) = -\frac{\hat{V}}{x^2N} + \Theta\left(\frac{1}{N^2}\right)
$$
the inequality \eqref{eq:Qcond1} can be written as
$$
\frac{\ln\hat{V}}{2N}-\frac{\hat{V}}{x^2N^2} +
\Theta\left(\frac{1}{N^3}\right)-\frac{x^2}{2\hat{V}}\ge \frac{\lambda\ln 2}{N}+\frac{\ln V}{2N}-\frac{x^2}{2V}
$$
That is,
$$
\frac{1}{V}-\frac{1}{\hat{V}}\ge \Theta\left(\frac{1}{N}\right)
$$
which leads to
$$
\hat{V}-V\ge \Theta\left(\frac{1}{N}\right)
$$
	
Now, we will find a sufficient condition for
\begin{equation}
2^\lambda Q\left(x\sqrt{\frac{N}{V}}\right)\ge Q\left(x\sqrt{\frac{N}{\hat{V}}}\right) \label{eq:Q2}
\end{equation}
The LHS of \eqref{eq:Qineqs} implies that
$$
2^\lambda Q\left(x\sqrt{\frac{N}{V}}\right)\ge \frac{2^\lambda\sqrt{V}}{x\sqrt{2\pi N}}\left(1-\frac{V}{x^2N}\right)e^{-\frac{x^2N}{2V}}
$$
while the RHS of \eqref{eq:Qineqs} implies that
$$
Q\left(x\sqrt{\frac{N}{\hat{V}}}\right)\le \frac{\sqrt{\hat{V}}}{x\sqrt{2\pi N}}e^{-\frac{x^2N}{2\hat{V}}}
$$
Therefore,
$$
\frac{\sqrt{\hat{V}}}{x\sqrt{2\pi N}}e^{-\frac{x^2N}{2\hat{V}}}\le \frac{2^\lambda\sqrt{V}}{x\sqrt{2\pi N}}\left(1-\frac{V}{x^2N}\right)e^{-\frac{x^2N}{2V}}
$$
is a sufficient condition for \eqref{eq:Q2}. This condition is equivalent to 
$$
\frac{\ln \hat{V}}{2} - \frac{x^2N}{2\hat{V}} \le
\frac{\ln V}{2} + \lambda\ln 2+\ln\left(1-\frac{V}{x^2N}\right)-\frac{x^2N}{2V}
$$
which means
$$
\frac{\ln\hat{V}}{2N}-\frac{x^2}{2\hat{V}} \le
\frac{\ln V}{2N} + \frac{\lambda\ln 2}{N} - \frac{V}{x^2N^2} +
\frac{V^2}{x^4}\Theta\left( \frac{1}{N^3} \right)-\frac{x^2}{2V}
$$
Therefore,
$$
\frac{1}{V}-\frac{1}{\hat{V}}\le\Theta\left(\frac{1}{N}\right)
$$
and
$$
\hat{V}-V\le\Theta\left(\frac{1}{N}\right)
$$
	
Summarizing, for some constants $\kappa_1$ and $\kappa_2$ (both independent of $N$) and $N$ sufficiently large, we have shown that if $\hat{V}_\lambda(N,W)-V_\lambda(W) \ge \kappa_1/N$ ($\hat{V}_\lambda(N,W)-V_\lambda(W) \le \kappa_2/N$, respectively) then the LHS of \eqref{eq:Qeq} is smaller (larger) than the RHS. In addition, note that the RHS of \eqref{eq:Qeq}, that defines $\hat{V}_\lambda(N,W)$, is monotonically increasing in $\hat{V}_\lambda(N,W)$. This proves our claim.
	
The proof for $V_{n,\lambda}(W)$ and $\hat{V}_{n,\lambda}(N,W)$ is identical to the proof for $V_{\lambda}(W)$ and $\hat{V}_{\lambda}(N,W)$, and is therefore omitted.
\hfill\(\IEEEQEDhere\)

\bibliographystyle{IEEEtran}
\bibliography{bibliography}
\end{document}

%% file: shortcuts.tex
\newcommand{\rem}[1]{}

\newtheorem{lemma}{Lemma}

\newtheorem{theorem}{Theorem}

\newcommand{\bre}{\begin{equation}}
\newcommand{\ere}{\end{equation}}

\newcommand{\ee}\]
\newcommand{\bra}{\begin{eqnarray}}
\newcommand{\era}{\end{eqnarray}}
\newcommand{\bfg}{\begin{figure}[hbtp]}
\newcommand{\efg}{\end{figure}}
\newcommand{\bver}{\begin{verbatim}}
\newcommand{\ever}{\end{verbatim}}
\newcommand{\bit}{\begin{itemize}}
\newcommand{\eit}{\end{itemize}}
\newcommand{\ben}{\begin{enumerate}}
\newcommand{\een}{\end{enumerate}}

\newcommand{\given}{\: | \:}

\newcommand{\baa}{\begin{eqnarray*}}
\newcommand{\eaa}{\end{eqnarray*}}

\newcommand{\bs}{{\bf s}}

\newcommand{\xor}{\oplus}

\newcommand{\cA}{{\cal A}}

\newcommand{\cX}{{\cal X}}
\newcommand{\cY}{{\cal Y}}

\newcommand{\cB}{{\cal B}}

\newcommand{\hR}{\hat{R}}

\newcommand{\defined}{\triangleq}

\def\argmax{\mathop{\rm argmax}}

\def\defined{\: {\stackrel{\scriptscriptstyle \Delta}{=}} \: }

\newfont{\boldlarge}{msbm10 scaled 1100}

\newcommand{\comment}[1]{}

\newlength{\tmpbigbar}


%% file: polarMAP030917.bbl
\begin{thebibliography}{10}
\providecommand{\url}[1]{#1}
\csname url@samestyle\endcsname
\providecommand{\newblock}{\relax}
\providecommand{\bibinfo}[2]{#2}
\providecommand{\BIBentrySTDinterwordspacing}{\spaceskip=0pt\relax}
\providecommand{\BIBentryALTinterwordstretchfactor}{4}
\providecommand{\BIBentryALTinterwordspacing}{\spaceskip=\fontdimen2\font plus
\BIBentryALTinterwordstretchfactor\fontdimen3\font minus
  \fontdimen4\font\relax}
\providecommand{\BIBforeignlanguage}[2]{{%
\expandafter\ifx\csname l@#1\endcsname\relax
\typeout{** WARNING: IEEEtran.bst: No hyphenation pattern has been}%
\typeout{** loaded for the language `#1'. Using the pattern for}%
\typeout{** the default language instead.}%
\else
\language=\csname l@#1\endcsname
\fi
#2}}
\providecommand{\BIBdecl}{\relax}
\BIBdecl

\bibitem{arikan2009channel}
E.~Arikan, ``{Channel polarization: A method for constructing
  capacity-achieving codes for symmetric binary-input memoryless channels},''
  \emph{IEEE Transactions on Information Theory}, vol.~55, no.~7, pp.
  3051--3073, 2009.

\bibitem{tal2015list}
I.~Tal and A.~Vardy, ``List decoding of polar codes,'' \emph{IEEE Transactions
  on Information Theory}, vol.~61, no.~5, pp. 2213--2226, 2015.

\bibitem{leroux2013semi}
C.~Leroux, A.~J. Raymond, G.~Sarkis, and W.~J. Gross, ``{A semi-parallel
  successive-cancellation decoder for polar codes},'' \emph{IEEE Transactions
  on Signal Processing}, vol.~61, no.~2, pp. 289--299, 2013.

\bibitem{li2014low}
B.~Li, H.~Shen, D.~Tse, and W.~Tong, ``{Low-latency polar codes via hybrid
  decoding},'' in \emph{Proc. 8th Int. Symp. Turbo Codes and Iterative Inf.
  Processing (ISTC)}, August 2014, pp. 223--227.

\bibitem{yuan2015low}
B.~Yuan and K.~Parhi, ``{Low-latency successive-cancellation list decoders for
  polar codes with multibit decision},'' \emph{IEEE Trans. Very Large Scale
  Integr. (VLSI) Syst.}, vol.~23, no.~10, pp. 2268--2280, 2015.

\bibitem{xiong2015symbol}
C.~Xiong, J.~Lin, and Z.~Yan, ``{Symbol-decision successive cancellation list
  decoder for polar codes},'' \emph{IEEE Transactions on Signal Processing},
  vol.~64, no.~3, pp. 675--687, February 2016.

\bibitem{mahdavifar2014performance}
H.~Mahdavifar, M.~El-Khamy, J.~Lee, and I.~Kang, ``Performance limits and
  practical decoding of interleaved reed-solomon polar concatenated codes,''
  \emph{Communications, IEEE Transactions on}, vol.~62, no.~5, pp. 1406--1417,
  May 2014.

\bibitem{trifonov2011generalized}
P.~Trifonov and P.~Semenov, ``{Generalized concatenated codes based on polar
  codes},'' in \emph{Proc. 8th International Symposium on Wireless
  Communication Systems (ISWCS)}, Nov 2011, pp. 442--446.

\bibitem{wang2016interleaved}
Y.~Wang, K.~R. Narayanan, and Y.~C. Huang, ``{Interleaved concatenations of
  polar codes with BCH and convolutional codes},'' \emph{IEEE Journal on
  Selected Areas in Communications}, vol.~34, no.~2, pp. 267--277, Feb 2016.

\bibitem{goldin2016block}
D.~Goldin and D.~Burshtein, ``{Block successive cancellation decoding of
  polarization-based codes},'' in \emph{Proc. 9th Int. Symp. Turbo Codes and
  Iterative Inf. Processing (ISTC)}, September 2016, pp. 256--260.

\bibitem{fossorier1995soft}
M.~P. Fossorier and S.~Lin, ``Soft-decision decoding of linear block codes
  based on ordered statistics,'' \emph{IEEE Transactions on Information
  Theory}, vol.~41, no.~5, pp. 1379--1396, 1995.

\bibitem{valembois2004box}
A.~Valembois and M.~Fossorier, ``Box and match techniques applied to
  soft-decision decoding,'' \emph{IEEE Transactions on Information Theory},
  vol.~50, no.~5, pp. 796--810, 2004.

\bibitem{nachmani2016learning}
E.~Nachmani, Y.~Be'ery, and D.~Burshtein, ``Learning to decode linear codes
  using deep learning,'' in \emph{Communication, Control, and Computing
  (Allerton), 2016 54th Annual Allerton Conference on}.\hskip 1em plus 0.5em
  minus 0.4em\relax IEEE, 2016, pp. 341--346.

\bibitem{nachmani2017deep}
E.~Nachmani, E.~Marciano, L.~Lugosch, W.~J. Gross, D.~Burshtein, and Y.~Beery,
  ``Deep learning methods for improved decoding of linear codes,'' \emph{arXiv
  preprint arXiv:1706.07043}, 2017.

\bibitem{tenbrink}
T.~Gruber, S.~Cammerer, J.~Hoydis, and S.~t. Brink, ``On deep learning-based
  channel decoding,'' \emph{accepted for CISS 2017, arXiv preprint
  arXiv:1701.07738}, 2017.

\bibitem{cammerer2017scaling}
S.~Cammerer, T.~Gruber, J.~Hoydis, and S.~t. Brink, ``Scaling deep
  learning-based decoding of polar codes via partitioning,'' \emph{arXiv
  preprint arXiv:1702.06901}, 2017.

\bibitem{polyanskiy2010channel}
Y.~Polyanskiy, H.~V. Poor, and S.~Verdu, ``Channel coding rate in the finite
  blocklength regime,'' \emph{IEEE Transactions on Information Theory},
  vol.~56, no.~5, pp. 2307--2359, May 2010.

\bibitem{korada2009polar}
S.~B. Korada, ``{Polar codes for channel and source coding},'' Ph.D.
  dissertation, EPFL, Lausanne, Switzerland, 2009.

\bibitem{mori2009per}
R.~Mori and T.~Tanaka, ``{Performance and construction of polar codes on
  symmetric binary-input memoryless channels},'' in \emph{Proc. IEEE
  International Symposium on Information Theory (ISIT)}, Seoul, Korea, June
  2009, pp. 1496 -- 1500.

\bibitem{ru_book}
T.~Richardson and R.~Urbanke, \emph{{Modern Coding Theory}}.\hskip 1em plus
  0.5em minus 0.4em\relax Cambridge, UK: Cambridge University Press, 2008.

\bibitem{galbook}
R.~G. Gallager, \emph{{Information Theory and Reliable Communication}}.\hskip
  1em plus 0.5em minus 0.4em\relax New York: Wiley, 1968.

\bibitem{barg2002random}
A.~Barg and G.~D. Forney, ``Random codes: minimum distances and error
  exponents,'' \emph{IEEE Transactions on Information Theory}, vol.~48, no.~9,
  pp. 2568--2573, Sep 2002.

\bibitem{alsan2014polarization}
M.~Alsan and E.~Telatar, ``{Polarization improves $E_0$},'' \emph{IEEE
  Transactions on Information Theory}, vol.~60, no.~5, pp. 2714--2719, May
  2014.

\bibitem{guillen2013extremes}
A.~G. i~Fabregas, I.~Land, and A.~Martinez, ``Extremes of error exponents,''
  \emph{IEEE Transactions on Information Theory}, vol.~59, no.~4, pp.
  2201--2207, April 2013.

\bibitem{erseghe2016coding}
T.~Erseghe, ``{Coding in the finite-blocklength regime: bounds based on Laplace
  integrals and their asymptotic approximations},'' \emph{IEEE Transactions on
  Information Theory}, vol.~62, no.~12, pp. 6854--6883, Dec 2016.

\bibitem{shannon1959probability}
C.~E. Shannon, ``Probability of error for optimal codes in a gaussian
  channel,'' \emph{The Bell System Technical Journal}, vol.~38, no.~3, pp.
  611--656, May 1959.

\bibitem{bch_matrices}
R.~H. Morelos-Zaragoza, \emph{The Art of Error Correcting Coding},
  2nd~ed.\hskip 1em plus 0.5em minus 0.4em\relax New York: Wiley, 2006, web
  site: \url{http://www.the-art-of-ecc.com}.

\bibitem{liva2016survey}
G.~Liva, L.~Gaudio, T.~Ninacs, and T.~Jerkovits, ``{Code design for short
  blocks: a survey},'' \emph{arXiv preprint arXiv:1610.00873}, 2016.

\bibitem{performance2016wonterghem}
J.~V. Wonterghem, A.~Alloumf, J.~J. Boutros, and M.~Moeneclaey, ``Performance
  comparison of short-length error-correcting codes,'' in \emph{2016 Symposium
  on Communications and Vehicular Technologies (SCVT)}, Nov 2016, pp. 1--6.

\end{thebibliography}
